\documentclass[aps,pra,showpacs,groupedaddress,superscriptaddress,twocolumn]{revtex4-1}
\usepackage{amssymb}
\usepackage{graphicx}
\usepackage{amsmath}

\newcommand{\bear}{\begin{eqnarray}}
\newcommand{\eear}{\end{eqnarray}}
\newcommand{\be}{\begin{equation}}
\newcommand{\ee}{\end{equation}}
\newcommand{\beqn}{\begin{eqnarray}}
\newcommand{\eeqn}{\end{eqnarray}}
\newcommand{\beqnn}{\begin{eqnarray*}}
\newcommand{\eeqnn}{\end{eqnarray*}}

\begin{document}

\title{Alguns aspectos da \'{O}ptica Qu\^{a}ntica usando campos luminosos em
modos viajantes \\ Some Aspects of Quantum Optics Using Light Fields in Traveling Waves}
\author{C.~Valverde}
\email{valverde@ueg.br}
\affiliation{C\^{a}mpus Henrique Santillo, Universidade
Estadual de Goi\'{a}s, BR 153, km 98, 75.132-903 An\'{a}polis, GO, Brazil} 
\affiliation{Universidade Paulista (UNIP), Rod. BR 153, km 7, 74.845-090
Goi\^ania, GO, Brazil}
\author{A. N. ~Castro}
\affiliation{C\^{a}mpus Henrique Santillo, Universidade
Estadual de Goi\'{a}s, BR 153, km 98, 75.132-903 An\'{a}polis, GO, Brazil} 
\affiliation{Universidade Paulista (UNIP), Rod. BR 153, km 7, 74.845-090
Goi\^ania, GO, Brazil}
\author{E. P. ~Santos}
\affiliation{C\^{a}mpus Henrique Santillo, Universidade
Estadual de Goi\'{a}s, BR 153, km 98, 75.132-903 An\'{a}polis, GO, Brazil}
\author{B.~Baseia}
\affiliation{Instituto de F\'{\i}sica, Universidade Federal de Goi\'as,
74001-970 Goi\^ania, GO, Brazil}
\date{\today }

\begin{abstract}
\textbf{Resumo:} Embora o tratamento te\'{o}rico para descrever o campo
luminoso na \'{O}ptica Qu\^{a}ntica fosse gen\'{e}rico, durante bom tempo
ele era referido predominantemente a modos \'{o}pticos aprisionados em
cavidades. Resultados importantes foram obtidos nesse cen\'{a}rio. Mas, em
vista da dificuldade pr\'{a}tica no uso desse campo, trazida pelos delet\'{e}%
rios efeitos de descoer\^{e}ncia sobre estados aprisionados, muitos f\'{\i}%
sicos da \'{a}rea passaram a focalizar com maior \^{e}nfase o tratamento em
modos viajantes. Neste breve relato tratamos o caso da engenharia de estados
n\~{a}o cl\'{a}ssicos da luz para mostrar alguns detalhes das aplica\c{c}%
\~{o}es nesse cen\'{a}rio. Aqui a intera\c{c}\~{a}o ``campo aprisionado-\'{a}%
tomo'', d\'{a} lugar \`{a} intera\c{c}\~{a}o ``campo viajante-separador de
feixes \'{o}pticos''. \newline

\textbf{Abstract:} Although the theoretical treatment to describe the light
field in Quantum Optics was generic, during large time it was predominantly
related to optical modes trapped inside cavities. Important results were
then obtained in this scenario. However, in view of the practical
difficulties due to the deleterious effects of decoherence upon states of
trapped fields, many physicists in this area began to focus more emphasis in
the treatment using traveling fields. This brief report concerns with
engineering non-classical states of light field to show some details and
applications in the later scenario. Here the interaction ``atom-trapped
field'', is translated to the interaction ``beam splitter-traveling field''.%
\newline
\end{abstract}

\maketitle

\section{Introdu\c{c}\~{a}o}

Tr\^{e}s anos ap\'{o}s a descoberta do laser, o fisico Roy J. Glauber
publicou em 1963 um importante trabalho te\'{o}rico onde tratava a luz
quanticamente \cite{glauber,gl1}. Seu artigo se tornou leitura obrigat\'{o}%
ria a pesquisadores da \'{a}rea da \'{O}ptica e correlatas; acrescido de
outros artigos afins tornou-se mais importante ainda com o resultado
experimental publicado em \cite{anti}: a descoberta do efeito
\textquotedblleft antiagrupamento de f\'{o}tons\textquotedblright\
(\textquotedblleft \textit{antibunching}\textquotedblright ) \cite%
{anti,a1,a2} que deu origem a uma nova \'{a}rea da F\'{\i}sica, a \'{O}ptica
Qu\^{a}ntica. Esse efeito constitui-se na primeira prova da exist\^{e}ncia
de efeito qu\^{a}ntico na radia\c{c}\~{a}o eletromagn\'{e}tica \cite{luz}.
Com efeito, a quantiza\c{c}\~{a}o dessa radia\c{c}\~{a}o j\'{a} tinha sido
feita em 1926, por Max Born, Werner Heisenberg e Ernst Pascual Jordan \cite%
{born}, inclusive sua aplica\c{c}\~{a}o em sistemas envolvendo intera\c{c}%
\~{a}o radia\c{c}\~{a}o-mat\'{e}ria, por Paul A. M. Dirac, em 1927 \cite%
{Dirac}. Mas faltava uma prova cabal da necessidade dessa quantiza\c{c}\~{a}%
o. \'{E} que, at\'{e} o resultado obtido em \cite{anti}, todos os efeitos
exibidos pela radia\c{c}\~{a}o eletromagn\'{e}tica podiam ser explicados
pela \'{O}ptica Cl\'{a}ssica, ou N\'{e}o-Cl\'{a}ssica \cite{Jaynes}; por
isso os f\'{\i}sicos n\~{a}o viam necessidade de uma teoria qu\^{a}ntica
para essa radia\c{c}\~{a}o. Estimulados pela descoberta desse primeiro
efeito qu\^{a}ntico, os f\'{\i}sicos da \'{a}rea passaram a investigar se a
radia\c{c}\~{a}o eletromagn\'{e}tica exibiria outros efeitos n\~{a}o explic%
\'{a}veis classicamente. Nessa busca, novos efeitos foram sendo encontrados.
Dentre eles citamos: (\textbf{i}) a estat\'{\i}stica sub-Poissoniana no
campo luminoso \cite{mandel,m1,m2,m3}; (\textbf{ii}) o efeito de compress%
\~{a}o do ruido do v\'{a}cuo qu\^{a}ntico (\textquotedblleft \textit{%
squeezing}\textquotedblright ) \cite{stoler,s1,s2,s3}; (\textbf{iii})\textit{%
\ }oscila\c{c}\~{o}es na invers\~{a}o at\^{o}mica \cite{schleich,cv1}; (%
\textbf{iv}) ocorr\^{e}ncia de zeros na distribui\c{c}\~{a}o de n\'{u}mero
de f\'{o}tons\textit{\ }\cite{zero}, etc.

Mas, qual era o crit\'{e}rio usado para definir um efeito \'{o}ptico como
genuinamente qu\^{a}ntico, isto \'{e}, sem explica\c{c}\~{a}o na \'{O}ptica
Cl\'{a}ssica? A identifica\c{c}\~{a}o do carater qu\^{a}ntico dos efeitos 
\'{o}pticos \'{e} feita descrevendo o estado do campo pelo operador
densidade $\hat{\rho}$ ( e n\~{a}o pela fun\c{c}\~{a}o de onda $\psi (\vec{r}%
,t)$\ ou pelo \textquotedblleft ket\textquotedblright\ $|\psi (t)\rangle $
); $\hat{\rho}$ \'{e} representado na base coerente inventada por Glauber: $%
\hat{\rho}=\int P(\alpha )|\alpha \rangle \langle \alpha |d^{2}\alpha $ e
verificam se a distribui\c{c}\~{a}o $P(\alpha )$ apresenta irregularidades,
tipo singularidades ou valores negativos \cite{glauber}. Se $P(\alpha )$
resultar assim intoler\'{a}vel, dizemos que o estado do campo \'{e} n\~{a}o
cl\'{a}ssico. \'{E} como ocorre no efeito t\'{u}nel: ele \'{e} qu\^{a}ntico
porque uma part\'{\i}cula cl\'{a}ssica s\'{o} tunelaria com velocidade imagin%
\'{a}ria - outra situa\c{c}\~{a}o intoler\'{a}vel. A componente $|\alpha
\rangle $\ \'{e} um dos estados coerentes da base coerente de Glauber; $%
|\alpha \rangle $\ \'{e} autovetor do operador $\hat{a}$: $\hat{a}|\alpha
\rangle =\lambda |\alpha \rangle ,$\ com $\lambda $\ $=\alpha $\ complexo,
pois o operador de aniquila\c{c}\~{a}o de f\'{o}tons, $\hat{a}|n\rangle \sim
|n-1\rangle ,$ n\~{a}o \'{e} hermitiano. Para uma compara\c{c}\~{a}o
ilustrativa, lembramos que um campo descrito por $\hat{\rho}$, representado
na base de n\'{u}mero (Fock), resultaria nessa outra forma: $\hat{\rho}=\sum
p_{n}|n\rangle \langle n|.$\ Nessa base teriamos $p_{n}=\frac{\langle \hat{n}%
\rangle ^{n}}{(1+\langle n\rangle )^{n+1}}$ se $\hat{\rho}$ fosse um campo
de luz t\'{e}rmica ou luz ca\'{o}tica de l\^{a}mpadas fluorescentes, ou
ainda luz de lasers funcionando abaixo do limiar, correspondendo \`{a}
distribui\c{c}\~{a}o de Bose-Einstein. Para esse campo luminoso a distribui%
\c{c}\~{a}o $P(\alpha )$ da base coerente de Glauber resulta em uma
Gaussiana, regular.

Al\'{e}m dos mencionados efeitos qu\^{a}nticos no campo de radia\c{c}\~{a}o,
foram observados tamb\'{e}m outros efeitos cujas explica\c{c}\~{o}es te\'{o}%
ricas s\'{o} funcionam se o campo eletromagn\'{e}tico \'{e} quantizado,
mesmo quando ele envolve o estado mais cl\'{a}ssico dentre os qu\^{a}nticos
- o estado coerente \cite{bb3,bb4}. Como exemplos citamos: (\textbf{a}) o
efeito colapso e ressurgimento (\textquotedblleft \textit{collapse-revival}%
\textquotedblright ) da invers\~{a}o at\^{o}mica, previsto teoricamente em
1980 por J. H. Eberly et al. \cite{Eberly} e observado em laborat\'{o}rio
por G. Rempe et al. \cite{Rempe}. Esse efeito ocorre quando \'{a}tomos de $2$
n\'{\i}veis interagem com convenientes campos eletromagn\'{e}ticos
quantizados \cite{Rempe,inv,inv1,cv3,cv4}; (\textbf{b})\emph{\ }espalhamento
de \'{a}tomos por luz em estado estacion\'{a}rio \cite{estacionarios,bbb}; (%
\textbf{c}) a superposi\c{c}\~{a}o de estados cl\'{a}ssicos do campo de radia%
\c{c}\~{a}o gerando um estado qu\^{a}ntico \cite{cv0,bb2}; um bem conhecido
exemplo \'{e} o chamado \textquotedblleft \textit{gato de Schr\"{o}dinger}%
\textquotedblright\ \cite{zurek,z0,z1,cv8}, proposto em 1935 por Erwin Schr%
\"{o}dinger para questionar fundamentos da Mec\^{a}nica Qu\^{a}ntica; (%
\textbf{d}) os estados do campo tendo correla\c{c}\~{a}o n\~{a}o local \cite%
{nlocal,n1,n2}, tamb\'{e}m chamados de estados emaranhados \textit{%
(\textquotedblleft entangled\textquotedblright ) }\cite%
{ent,e1,e2,e3,e4,e5,e6,e7,e8}, utilizados nos processos de teletransporte qu%
\^{a}ntico \cite{tele,t1,t2}, importantes em aplica\c{c}\~{o}es na
engenharia de estados qu\^{a}nticos \cite%
{cv0,eng,eng1,eng2,eng3,eng4,eng5,cv6,eg1,eg2,eg3,eg4,eg5,eg6,eg7}, tanto no
caso de campos estacion\'{a}rios em excelentes cavidades eletromagn\'{e}%
ticas \cite{cav} (para evitar descoer\^{e}ncia dos estados), como no caso de
campos viajantes, atravessando arranjos \'{o}pticos como fontes de feixes
luminosos, espelhos, prismas, divisores de feixes (\textquotedblleft \textit{%
beam splitters}\textquotedblright ) \cite{beam,orszag,cv2} e detectores de f%
\'{o}tons. Tais estados s\~{a}o tamb\'{e}m \'{u}teis na computa\c{c}\~{a}o qu%
\^{a}ntica \cite{cq,bb1,cv5,cq1,cv7}, na criptografia qu\^{a}ntica \cite%
{crq,crq1,crq2} e no teletransporte, tanto de estados at\^{o}micos como de
estados de campos de radia\c{c}\~{a}o eletromagn\'{e}tica \cite%
{tq,tq1,tq2,e5,e7}.

Neste trabalho vamos considerar o caso de modos viajantes, incidindo em um
arranjo \'{o}ptico contendo $1$ ou $2$ separadores de feixes, espelhos e
detectores de f\'{o}tons. O objetivo \'{e} mostrar com detalhes, e para cada
estado do campo luminoso incidente na entrada do arranjo \'{o}ptico, em que
tipo de estado o campo emerge na sa\'{\i}da do mesmo. A finalidade \'{e}
engenheirar estados \'{o}pticos n\~{a}o cl\'{a}ssicos que apresentem alguma
propriedade aplicativa. Sabe-se que a pesquisa usando estados \'{o}pticos de
modos viajantes tem sido ultimamente mais explorada que a pesquisa usa
estados \'{o}pticos em cavidades, a raz\~{a}o sendo a menor sensibilidade
dos modos viajantes a delet\'{e}rios efeitos de descoer\^{e}ncia.
Aproveitando essa qualidade e tamb\'{e}m a vantagem dada pelo baixo custo
dos componentes em arranjos \'{o}pticos usados em modos viajantes, muitos
grupos de pesquisa experimental, brasileiros, da \'{a}rea \'{O}ptica Qu\^{a}%
ntica, t\^{e}m concentrado seus esfor\c{c}os nessa linha.

\section{Separador de feixes e suas propriedades}

\begin{figure}[tbh]
\begin{center}
%\vspace{-0.5cm}
\includegraphics[width=0.40\textwidth]{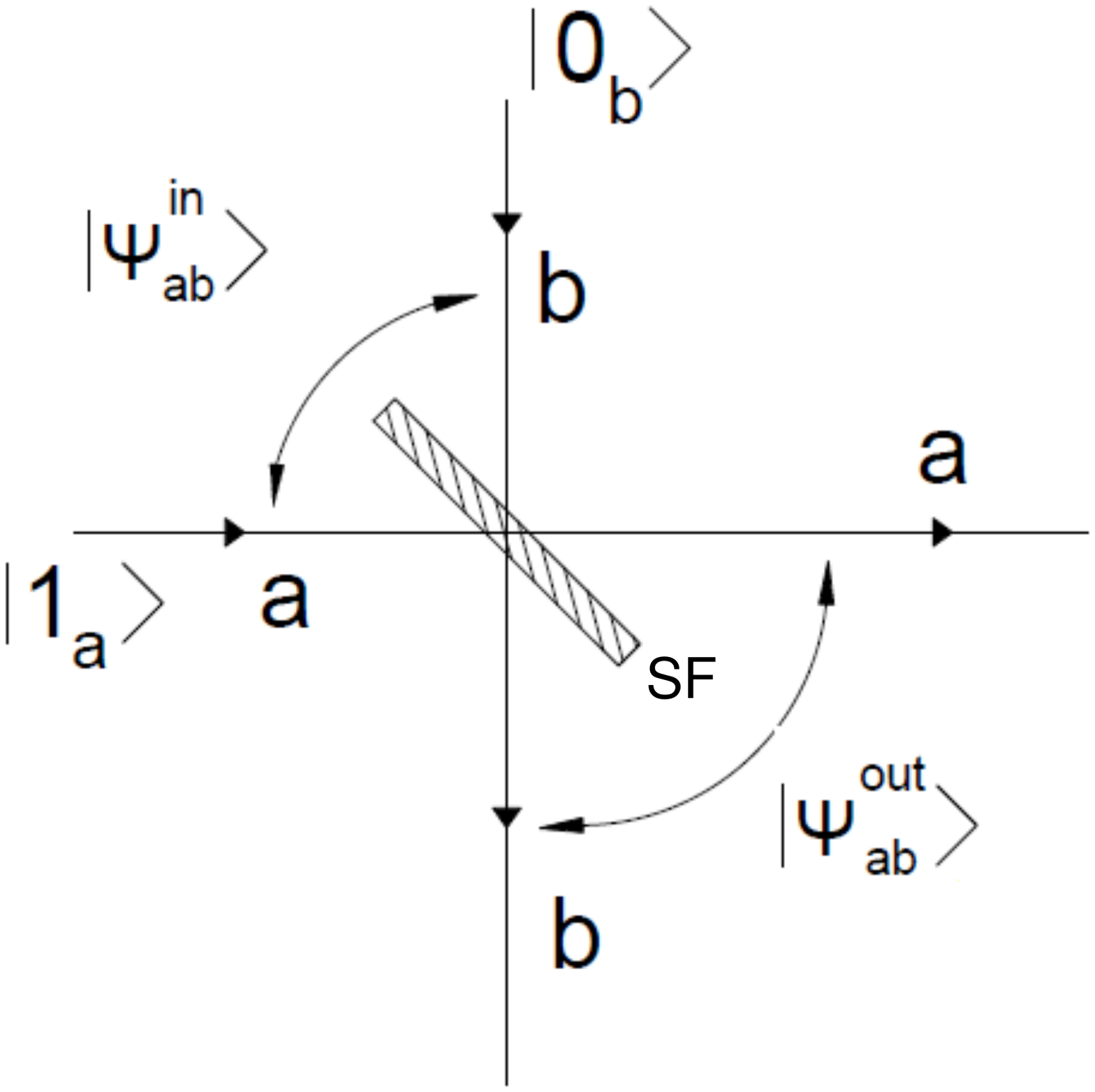}
\end{center}
\par
%\vspace{-1cm}
\caption{Luz incidindo e emergindo em dois modos de um separador de feixes (%
\textbf{SF)}. }
\label{b1}
\end{figure}

No caso da radia\c{c}\~{a}o luminosa tratada quanticamente, quando ela
incide em um separador de feixes ($SF$) como mostrado na Fig.(\ref{b1}), na
saida desse dispositivo \'{o}ptico o feixe no estado \textit{total} $%
\left\vert \psi (\tau )\right\rangle $ emerge representado pela express\~{a}%
o,%
%________________________FIGURA 43.1 _________________________________________
\begin{equation}
\left\vert \psi (\tau )\right\rangle =\exp ^{-\frac{i}{\hslash }\tau \hat{H}%
}\left\vert \psi (0)\right\rangle ,  \label{1a}
\end{equation}%
onde $\hat{H}=\hslash \lambda (\hat{a}\hat{b}^{\dagger }+\hat{a}^{\dagger }%
\hat{b})$ \'{e} o operador hamiltoniano e representa a a\c{c}\~{a}o do $SF$
sobre o feixe de luz; $\lambda $ \'{e} intensidade da intera\c{c}\~{a}o
entre os modos $\mathbf{a}$ e $\mathbf{b,}$\textbf{\ }ver Fig.(\ref{b1}), $%
\tau $ \'{e} o tempo de travessia no dispositivo $SF$ e $\left\vert \psi
(0)\right\rangle =\left\vert \psi (\tau =0)\right\rangle $ representa o
estado inicial do feixe \textit{total} incidindo pelos canais $\mathbf{a}$ e 
$\mathbf{b}$ do $SF.$ Podemos escrever a Eq. (\ref{1a}) na forma, 
\begin{equation}
\left\vert \psi (\tau )\right\rangle =e^{-i\lambda \tau (\hat{a}\hat{b}%
^{\dagger }+\hat{a}^{\dagger }\hat{b})}\text{ }\left\vert \psi
(0)\right\rangle .
\end{equation}%
Anotaremos doravante o estado inicial incidente $\left\vert \psi
(0)\right\rangle =\left\vert \psi _{ab}^{in}\right\rangle $ e o estado
emergente $\left\vert \psi (\tau )\right\rangle =\left\vert \psi
_{ab}^{out}\right\rangle $. O termo \textquotedblleft in\textquotedblright\ 
\'{e} para incidente (\textit{ingoing}) e \textquotedblleft
out\textquotedblright\ \'{e} para de emergente (\textit{outgoing}). Temos ent%
\~{a}o,%
\begin{equation}
\left\vert \psi _{ab}^{out}\right\rangle =e^{-i\lambda \tau (\hat{a}\hat{b}%
^{\dagger }+\hat{a}^{\dagger }\hat{b})}\text{ }\left\vert \psi
_{ab}^{in}\right\rangle .  \label{1a1}
\end{equation}%
\qquad \qquad

Tendo em vista a Fig.(\ref{b1}) e visando simplificar os c\'{a}lculos, conv%
\'{e}m reescrever a equa\c{c}\~{a}o acima na forma,%
\begin{equation}
\left\vert \psi _{ab}^{out}\right\rangle =\hat{S}_{ab}\left\vert \psi
_{ab}^{in}\right\rangle ,  \label{1b}
\end{equation}%
onde $\hat{S}_{ab}=e^{-i\lambda \tau (\hat{a}\hat{b}^{\dagger }+\hat{a}%
^{\dagger }\hat{b})}$ \'{e} um operador unit\'{a}rio, isto \'{e}, $\hat{S}%
_{ab}^{\dag }\hat{S}_{ab}=\hat{S}_{ab}\hat{S}_{ab}^{\dagger }=\mathbf{\hat{1}%
}$. A unitariedade de $\hat{S}_{ab}$ permite obter esta \'{u}til express\~{a}%
o, 
\begin{equation}
\hat{S}_{ab}\hat{a}^{\dagger }\hat{S}_{ab}^{\dagger }=T\hat{a}^{\dagger }+iR%
\hat{b}^{\dagger },  \label{1b1}
\end{equation}%
onde $T=\cos (\lambda \tau )$ e $R=\sin (\lambda \tau )$ representam
respectivamente os coeficientes de transmiss\~{a}o e reflex\~{a}o do $SF$.

Para provar o resultado da Eq. (\ref{1b1}) usaremos uma identidade muito
conhecida, do cap\'{\i}tulo sobre \'{a}lgebra de operadores nos livros de Mec%
\^{a}nica Qu\^{a}ntica e/ou \'{O}ptica Qu\^{a}ntica \cite%
{luz,Dirac,Jaynes,zero,cav,orszag,louisell},

\begin{widetext}
\begin{equation}
e^{i\chi \hat{A}}\hat{B}e^{-i\chi \hat{A}}=\hat{B}+(i\chi )[\hat{A},\hat{B}]+%
\frac{(i\chi )^{2}}{2!}[\hat{A},[\hat{A},\hat{B}]]+\frac{(i\chi )^{3}}{3!}[%
\hat{A},[\hat{A},[\hat{A},\hat{B}]]]+...\text{ },  \label{1b2}
\end{equation}%
na qual, fazendo $\chi =\lambda \tau $, $\ \hat{A}=(\hat{a}^{\dagger }\hat{b}%
+\hat{a}\hat{b}^{\dagger })$, $\hat{B}=\hat{a}^{\dagger }$\ bem como usando
os comutadores $[\hat{a},\hat{a}^{\dagger }]=1$ e $[\hat{a},\hat{b}]=0$
obtemos,

\begin{eqnarray}
\lbrack \hat{A},\hat{B}] &=&[(\hat{a}^{\dagger }\hat{b}+\hat{a}\hat{b}%
^{\dagger }),\hat{a}^{\dagger }]  \notag \\
&=&(\hat{a}^{\dagger }\hat{b}+\hat{a}\hat{b}^{\dagger })\hat{a}^{\dagger }-%
\hat{a}^{\dagger }(\hat{a}^{\dagger }\hat{b}+\hat{a}\hat{b}^{\dagger })
\notag \\
&=&\hat{a}^{\dagger }\hat{a}^{\dagger }\hat{b}+\hat{a}\hat{a}^{\dagger }\hat{%
b}^{\dagger }-\hat{a}^{\dagger }\hat{a}^{\dagger }\hat{b}-\hat{a}^{\dagger }%
\hat{a}\hat{b}^{\dagger }  \notag \\
&=&\hat{b}^{\dagger },  \label{1b3}
\end{eqnarray}%
\end{widetext}
enquanto que teremos $[\hat{A},[\hat{A},\hat{B}]]=[(\hat{a}^{\dagger }\hat{b}%
+\hat{a}\hat{b}^{\dagger }),\hat{b}^{\dagger }]=\hat{a}^{\dagger }$ e depois
teremos $[\hat{A},[\hat{A},[\hat{A},\hat{B}]]]=$\ $[(\hat{a}^{\dagger }\hat{b%
}+\hat{a}\hat{b}^{\dagger }),a^{\dagger }]=\hat{b}^{\dagger },...$\ etc.

Esse resultado mostra que a sequ\^{e}ncia de comutadores da express\~{a}o
Eq. (\ref{1b2}) resulta nesta outra sequ\^{e}ncia: $\hat{a}^{\dagger },$ $%
b^{\dagger },$ $\hat{a}^{\dagger },$ $b^{\dagger },$ $\hat{a}^{\dagger },$ $%
\hat{b}^{\dagger },...$ \ acompanhada de pot\^{e}ncias crescentes na forma $%
\pm \frac{(i\chi )^{n}}{n!}.$ De modo que, nas pot\^{e}ncias \'{\i}mpares de 
$(i\chi )$ os comutadores da Eq. (\ref{1b2}) s\~{a}o do tipo $\pm \hat{b}%
^{\dagger }$ enquanto nas pot\^{e}ncias pares os comutadores s\~{a}o do tipo 
$\pm \hat{a}^{\dagger }$. Assim, denotando na Eq. (\ref{1b2}) $e^{i\chi \hat{%
A}}=\hat{S}_{ab}$\ e $\hat{B}=\hat{a}^{\dagger }$ obtemos, usando os j\'{a}
mencionados $\hat{S}_{ab}=e^{-i\chi (\hat{a}\hat{b}^{\dagger }+\hat{a}%
^{\dagger }\hat{b})}$ e $\chi =\lambda \tau ,$

\begin{widetext}
\begin{equation}
\hat{S}_{ab}\hat{a}^{\dagger }\hat{S}_{ab}^{\dag }=[1+\frac{(i\chi )^{2}}{2!}%
+\frac{(i\chi )^{4}}{4!}+\frac{(i\chi )^{6}}{6!}+...]\hat{a}^{\dagger
}+[(i\chi )+\frac{(i\chi )^{3}}{3!}+\frac{(i\chi )^{5}}{5!}+\text{ }...]\hat{%
b}^{\dagger },  \label{2b}
\end{equation}%
que pode ser reescrita na forma,

\begin{equation}
\hat{S}_{ab}\hat{a}^{\dagger }\hat{S}_{ab}^{\dag }=(1-\frac{\chi ^{2}}{2!}+%
\frac{\chi ^{4}}{4!}-\frac{\chi ^{6}}{6!}+...)\hat{a}^{\dagger }+i(\chi -%
\frac{\chi ^{3}}{3!}+\frac{\chi ^{5}}{5!}-\text{ }...)\hat{b}^{\dagger },
\label{2b0}
\end{equation}%
e compactada assim,

\begin{equation}
\hat{S}_{ab}\hat{a}^{\dagger }\hat{S}_{ab}^{\dag }=\sum\limits_{n=0}^{\infty
}\frac{(-1)^{n}}{(2n)!}\chi ^{2n}\hat{a}^{\dagger
}+i\sum\limits_{n=0}^{\infty }\frac{(-1)^{n}}{(2n+1)!}\chi ^{2n+1}\hat{b}%
^{\dagger },  \label{2b00}
\end{equation}%
\end{widetext}
onde reconhecemos as duas s\'{e}ries de pot\^{e}ncias: $\sum\limits_{n=0}^{%
\infty }\frac{(-1)^{n}}{(2n)!}\chi ^{2n}=\cos (\chi )=T$ \ e $%
\sum\limits_{n=0}^{\infty }\frac{(-1)^{n}}{(2n+1)!}\chi ^{2n+1}=\sin (\chi
)=R$. Isto permite escrever a Eq. (\ref{2b00}) na forma final, 
\begin{eqnarray}
\hat{S}_{ab}\hat{a}^{\dagger }S_{ab}^{\dag } &=&\cos (\chi )\hat{a}^{\dagger
}+i\sin (\chi )\hat{b}^{\dagger }  \notag \\
&=&T\hat{a}^{\dagger }+iR\hat{b}^{\dagger }.  \label{2b1}
\end{eqnarray}%
\qquad

Um procedimento an\'{a}logo aplicado ao operador $\hat{b}^{\dagger }$
fornece,%
\begin{equation}
\hat{S}_{ab}\hat{b}^{\dagger }S_{ab}^{\dag }=T\hat{b}^{\dagger }+iR\hat{a}%
^{\dagger }.  \label{2bb}
\end{equation}

As Eqs. (\ref{2b1}), (\ref{2bb}) s\~{a}o b\'{a}sicas para o que segue.

\section{Aplica\c{c}\~{o}es do separador de feixe}

\subsection{Caso 1:}

Como aplica\c{c}\~{a}o inicial e preliminar para os demais t\'{o}picos da
sequ\^{e}ncia, vamos considerar o caso mais simples, mostrado na Fig.(\ref%
{b1}), em que o estado \textit{total} do feixe de luz na entrada do $SF$
corresponde \`{a} express\~{a}o matem\'{a}tica operacional seguinte,
representando o estado $\left\vert 1\right\rangle $, de um f\'{o}ton
incidente no modo \textbf{a} do $SF$, e o estado $\left\vert 0\right\rangle $%
, de zero f\'{o}ton incidente no modo \textbf{b}, 
\begin{equation}
\left\vert \psi _{ab}^{in}\right\rangle =\left\vert 1\right\rangle
_{a}\left\vert 0\right\rangle _{b}.  \label{2b2}
\end{equation}%
O estado do feixe luminoso que emerge do $SF$ \'{e} obtido assim, passo a
passo, 
\begin{eqnarray}
\left\vert \psi _{ab}^{out}\right\rangle &=&\hat{S}_{ab}\left\vert \psi
_{ab}^{in}\right\rangle ,  \notag \\
&=&\hat{S}_{ab}\left\vert 1\right\rangle _{a}\left\vert 0\right\rangle _{b},
\notag \\
&=&\hat{S}_{ab}(\hat{a}^{\dagger }\left\vert 0\right\rangle _{a})\left\vert
0\right\rangle _{b},  \notag \\
&=&\hat{S}_{ab}\hat{a}^{\dagger }(\hat{S}_{ab}^{\dagger }\hat{S}%
_{ab})\left\vert 0\right\rangle _{a}\left\vert 0\right\rangle _{b},  \notag
\\
&=&(\hat{S}_{ab}\hat{a}^{\dagger }\hat{S}_{ab}^{\dagger })\hat{S}%
_{ab})\left\vert 0\right\rangle _{a}\left\vert 0\right\rangle _{b},  \notag
\\
&=&(Ta^{\dagger }+iRb^{\dagger })\left\vert 0\right\rangle _{a}\left\vert
0\right\rangle _{b},  \notag \\
&=&T\left\vert 1\right\rangle _{a}\left\vert 0\right\rangle
_{b}+iR\left\vert 0\right\rangle _{a}\left\vert 1\right\rangle _{b}.
\label{2b3}
\end{eqnarray}

O procedimento alg\'{e}brico que levou a esse resultado ser\'{a} estendido
abaixo.

\subsection{Caso 2:}

\begin{figure}[tbh]
\begin{center}
%\vspace{-0.5cm}
\includegraphics[width=0.40\textwidth]{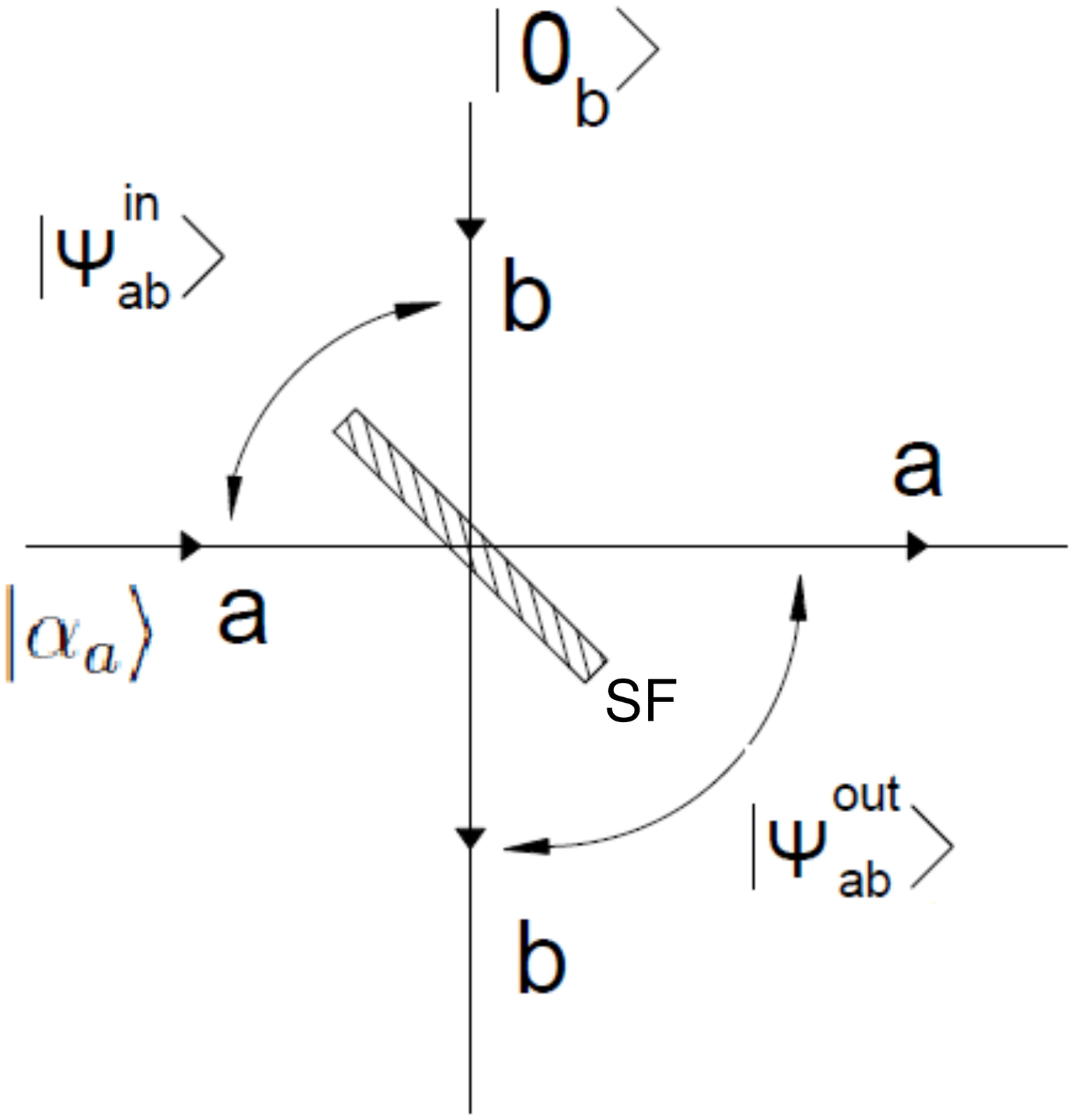}
\end{center}
\par
%\vspace{-1cm}
\caption{O mesmo que na Fig.(\protect\ref{b1}): um dos feixes no estado
coerente $\left\vert \protect\alpha \right\rangle $ incide no modo $\mathbf{%
a,}$ outro no estado de v\'{a}cuo $\left\vert 0\right\rangle ,$ no modo $%
\mathbf{b}$. }
\label{b2}
\end{figure}

Separadores de feixe ($SFs$) modificam os estados do campo luminoso e
portanto mudam tamb\'{e}m suas estat\'{\i}sticas, com excess\~{a}o dos
estados coerentes, $|\alpha \rangle $. Para prov\'{a}-lo partimos da Fig.(%
\ref{b2}).

%________________________FIGURA 43.1 _________________________________________

No modo $\mathbf{a,}$ temos um feixe incidente, no estado coerente $|\alpha
\rangle _{a}=\hat{D}_{a}(\alpha )\left\vert 0\right\rangle _{a}$\ - o mais cl%
\'{a}ssico dentre os qu\^{a}nticos\cite{barry}; $\hat{D}_{a}(\alpha
)=e^{(\alpha \hat{a}\dag -\alpha ^{\ast }\hat{a})}$. Nesse caso, usando
novamente o estado de v\'{a}cuo $\left\vert 0\right\rangle _{b}$ no modo $%
\mathbf{b}$, os passos para chegar ao resultado buscado s\~{a}o esses, 
\begin{eqnarray}
|\psi _{ab}^{out}\rangle &=&\hat{S}_{ab}|\alpha \rangle _{a}\left\vert
0\right\rangle _{b}  \notag \\
&=&\hat{S}_{ab}\hat{D}_{a}(\alpha )\left\vert 0\right\rangle _{a}\left\vert
0\right\rangle _{b},  \notag \\
&=&\hat{D}_{a}(T\alpha )\left\vert 0\right\rangle _{a}\hat{D}_{b}(iR\alpha
)\left\vert 0\right\rangle _{b},  \notag \\
&=&|T\alpha \rangle _{a}|iR\alpha \rangle _{b}.  \label{3c3}
\end{eqnarray}%
Esse resultado mostra que os f\'{o}tons do feixe luminoso,\ distribuem-se
aos dois feixes que emergem nos estados coerentes, $|T\alpha \rangle _{a}$\
e $|iR\alpha \rangle _{b},$\ dos modos $\mathbf{a}$\ e $\mathbf{b}$\ do $SF$%
. N\~{a}o h\'{a} ganho nem perda de energia, pois $T\in \lbrack 0,1]$, $R\in
\lbrack 0,1]$\ com $T^{2}+R^{2}=1$\ para o suposto $SF$ ideal: o n\'{u}mero 
\textit{total} de f\'{o}tons na entrada do $SF$ \'{e} o mesmo do \textit{%
total} na saida. De fato, o n\'{u}mero m\'{e}dio de f\'{o}tons emergindo nos
modos $\mathbf{a}$\ e $\mathbf{b}$\ resulta respectivamente $\langle
n\rangle _{a}=|T\alpha |^{2}$\ e $\langle \hat{n}\rangle _{b}=|iR\alpha
|^{2} $, cuja soma \'{e} $|\alpha |^{2}(T^{2}+R^{2})=|\alpha |^{2}$, o mesmo
valor m\'{e}dio no `input': $|\alpha |^{2}+0^{2}=|\alpha |^{2}$\ \cite%
{bb,bblivro}. Notamos tamb\'{e}m que os estados de entrada no $SF,$\ $%
|\alpha \rangle _{a}$\ e $|0\rangle $\ s\~{a}o estados coerentes, pois o
estado de v\'{a}cuo \'{e} um estado coerente trivial, ele satisfaz \`{a}
defini\c{c}\~{a}o: $\hat{a}|\alpha \rangle =\alpha |\alpha \rangle $ pois $%
\hat{a}|0\rangle =0|0\rangle .$ Tamb\'{e}m, os estados na sa\'{\i}da, $%
|T\alpha \rangle _{a}$\ e $|iR\alpha \rangle _{b},$\ s\~{a}o coerentes pois $%
R$ e $T$ s\~{a}o n\'{u}meros. Logo, o $SF$\ n\~{a}o mudou a estat\'{\i}stica
dos estados, uma vez que todos os estados coerentes exibem estat\'{\i}stica
poissoniana.

Agora provaremos o resultado na Eq. (\ref{3c3}); vamos denotar os operadores 
$\hat{a},$ de aniquila\c{c}\~{a}o de f\'{o}tons nos campos incidentes, por $%
\hat{a}_{in}.$ Os resultados das Eqs. (\ref{2b1}) e (\ref{2bb}) permitem
relacionar o operador $\hat{a}_{in}$ com o operador $\hat{a}_{out},$ de
aniquila\c{c}\~{a}o de f\'{o}tons nos campos emergentes, atrav\'{e}s da
representa\c{c}\~{a}o matricial usada em problemas de espalhamento \cite%
{orszag}, 
\begin{equation}
\left( 
\begin{array}{c}
\hat{a}_{out} \\ 
\hat{b}_{out}%
\end{array}%
\right) =\left( 
\begin{array}{cc}
T & iR \\ 
iR & T%
\end{array}%
\right) \left( 
\begin{array}{c}
\hat{a}_{in} \\ 
\hat{b}_{in}%
\end{array}%
\right) ,  \label{4c3}
\end{equation}%
onde $T^{2}+R^{2}=1.$ Da equa\c{c}\~{a}o acima resulta que $\hat{a}_{out}=T%
\hat{a}_{in}+iR\hat{b}_{in}$ e $\hat{b}_{out}=iR\hat{a}_{in}+T\hat{b}_{in}$.
Revertendo o sistema matricial acima encontramos,

\begin{equation}
\left( 
\begin{array}{c}
\hat{a}_{in} \\ 
\hat{b}_{in}%
\end{array}%
\right) =\left( 
\begin{array}{cc}
T & -iR \\ 
-iR & T%
\end{array}%
\right) \left( 
\begin{array}{c}
\hat{a}_{out} \\ 
\hat{b}_{out}%
\end{array}%
\right)  \label{4c4}
\end{equation}%
ou, equivalentemente,

\begin{eqnarray}
\hat{a}_{in} &=&T\hat{a}_{out}-iR\hat{b}_{out}\text{ },  \label{4c5} \\
\hat{b}_{in} &=&-iR\hat{a}_{out}+T\hat{b}_{out}  \label{4c6}
\end{eqnarray}%
que permite escrever, com \ $\hat{D}_{a}(\alpha )=e^{(\alpha \hat{a}%
_{in}^{\dagger }-\alpha ^{\ast }\hat{a}_{in})},$%
\begin{eqnarray}
|\psi _{ab}^{in}\rangle &=&|\alpha \rangle _{a}\left\vert 0\right\rangle
_{b}=\hat{D}_{a}(\alpha )\left\vert 0\right\rangle _{a}\left\vert
0\right\rangle _{b}\text{ },  \notag \\
&=&e^{(\alpha \hat{a}_{in}^{\dagger }-\alpha ^{\ast }\hat{a}%
_{in})}\left\vert 0\right\rangle _{a}\left\vert 0\right\rangle _{b}\text{ }.
\label{4c7}
\end{eqnarray}%
Usando a Eq. (\ref{4c5}) e sua adjunta, mudamos a Eq. (\ref{4c7}) para a
forma,%
\begin{equation}
|\psi _{ab}^{out}\rangle =e^{[\alpha (T\hat{a}_{out}^{\dagger }+iR\hat{b}%
_{out}^{\dagger })-\alpha ^{\ast }(T\hat{a}_{out}-iR\hat{b}%
_{out})]}\left\vert 0\right\rangle _{a}\left\vert 0\right\rangle _{b},
\label{4c8}
\end{equation}%
que pode ser reescrita como,%
\begin{eqnarray}
|\psi _{ab}^{out}\rangle &=&e^{(\alpha T\hat{a}_{out}^{\dagger }-\alpha
^{\ast }T\hat{a}_{out})}\text{ }e^{(iR\alpha \hat{b}_{out}^{\dagger }-\alpha
^{\ast }(iR)^{\ast }\hat{b}_{out})}\left\vert 0\right\rangle _{a}\left\vert
0\right\rangle _{b}\text{ },  \notag \\
&=&\hat{D}_{a}(T\alpha )\hat{D}_{b}(iR\alpha )\left\vert 0\right\rangle
_{a}\left\vert 0\right\rangle _{b},  \notag \\
&=&(\hat{D}_{a}(T\alpha )\left\vert 0\right\rangle _{a})(\hat{D}%
_{b}(iR\alpha )\left\vert 0\right\rangle _{b}),  \notag \\
&=&|T\alpha \rangle _{a}|iR\alpha \rangle _{b}.  \label{4c9}
\end{eqnarray}%
que prova o resultado antecipado na Eq. (\ref{3c3}). Para interpretar o
estado $|T\alpha \rangle _{a}$ como coerente usamos a analogia: se\ $%
D_{a}(\alpha )|0\rangle _{a}=|\alpha \rangle _{a}$\ ent\~{a}o $D_{a}(T\alpha
)|0\rangle _{a}=|T\alpha \rangle _{a}$. Ambos, $|\alpha \rangle _{a}$\ e $%
|T\alpha \rangle _{a\text{ }},$\ s\~{a}o estados coerentes, gerados pela atua%
\c{c}\~{a}o dos operadores $D_{a}(\alpha )$\ e $D_{a}(T\alpha )$\ no estado $%
|0\rangle _{a}$ respectivamente. O mesmo vale para o estado $|iR\alpha
\rangle _{b}$.

\subsection{Caso 3:}

Vamos agora considerar dois feixes em estados gen\'{e}ricos, incidindo nos
modos $\mathbf{a}$ e $\mathbf{b}$ do $SF$, conforme mostrado na Fig.(\ref%
{3bb}).

%________________________FIGURA 43.1 _________________________________________
\begin{figure}[tbh]
\begin{center}
%\vspace{-0.5cm}
\includegraphics[width=0.40\textwidth]{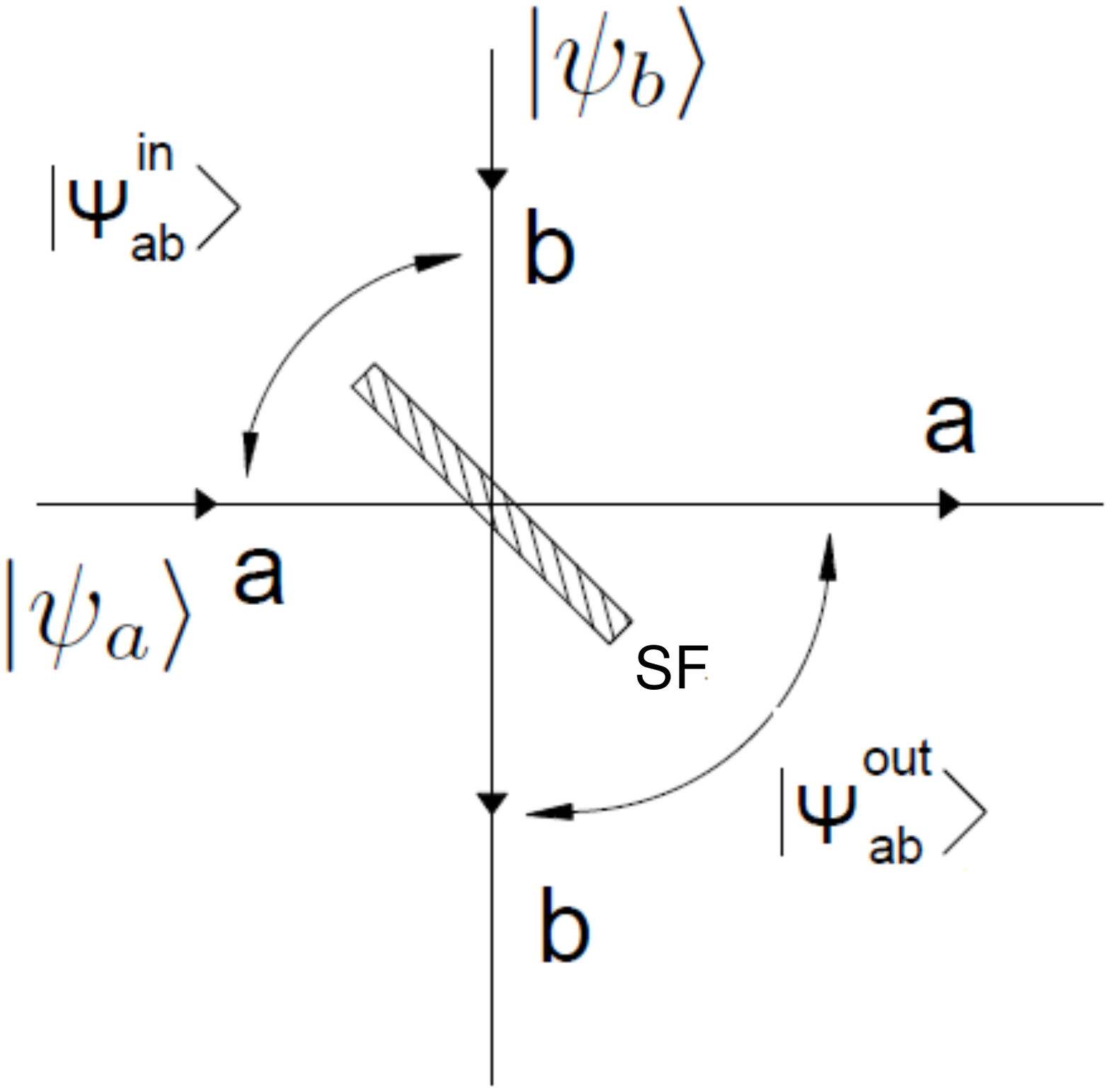}
\end{center}
\par
%\vspace{-1cm}
\caption{O mesmo que nas Figs.(\protect\ref{b1} e \protect\ref{b2}), para
incid\^{e}ncias de feixes nos estados $\left\vert \protect\psi %
_{a}\right\rangle ,$ no modo $\mathbf{a}$ e $\left\vert \protect\psi %
_{b}\right\rangle ,$ modo $\mathbf{b}$. }
\label{3bb}
\end{figure}
Temos,

\begin{equation}
\left\vert \psi _{ab}^{out}\right\rangle =\hat{S}_{ab}\left\vert \psi
_{ab}^{in}\right\rangle =\hat{S}_{ab}\left\vert \psi _{a}\right\rangle
\left\vert \psi _{b}\right\rangle ,  \label{4c10}
\end{equation}%
onde expandiremos na base de Fock: $\left\vert \psi _{a}\right\rangle
=\sum\limits_{n}C_{n}\left\vert n\right\rangle $ e $\left\vert \psi
_{b}\right\rangle =\sum\limits_{m}C_{m}\left\vert m\right\rangle ,$ para
obter,%
\begin{eqnarray}
|\psi _{ab}^{out}\rangle &=&\hat{S}_{ab}(\sum\limits_{n}C_{n}|n\rangle _{a})%
\text{ }(\sum\limits_{m}C_{m}\left\vert m\right\rangle _{b})  \notag \\
&=&\hat{S}_{ab}\sum\limits_{n}C_{n}\sum\limits_{m}C_{m}|n\rangle
_{a}\left\vert m\right\rangle _{b},  \label{4d}
\end{eqnarray}%
e agora usando $|n\rangle _{a}=$ $\frac{(\hat{a}^{\dagger })^{n}}{\sqrt{n!}}%
|0\rangle _{a}$\ e $|m\rangle _{b}=$ $\frac{(b^{\dagger })m}{\sqrt{m!}}%
|0\rangle _{b}$ resulta,

\begin{eqnarray}
\left\vert \psi _{ab}^{out}\right\rangle &=&\hat{S}_{ab}\sum\limits_{n}C_{n}%
\frac{(\hat{a}^{\dagger })^{n}}{\sqrt{n!}}\sum\limits_{m}C_{m}\frac{%
(b^{\dagger })^{m}}{\sqrt{m!}}\left\vert 0\right\rangle _{a}\left\vert
0\right\rangle _{b},  \notag \\
&=&\sum\limits_{n}C_{n}\frac{\hat{S}_{ab}(\hat{a}^{\dagger })^{n}\mathbf{%
\hat{1}}}{\sqrt{n!}}\sum\limits_{m}C_{m}\frac{(b^{\dagger })^{m}\mathbf{\hat{%
1}}}{\sqrt{m!}}(\left\vert 0\right\rangle _{a}\left\vert 0\right\rangle
_{b}),  \notag \\
&=&\sum\limits_{n,m}C_{n},m\frac{\hat{S}_{ab}(\hat{a}^{\dagger })^{n}\mathbf{%
\hat{1}}}{\sqrt{n!}}\frac{(b^{\dagger })^{m}\mathbf{\hat{1}}}{\sqrt{m!}}%
(\left\vert 0\right\rangle _{a}\left\vert 0\right\rangle _{b}).  \label{4e}
\end{eqnarray}%
Nos passos acima inserimos o operador $\mathbf{\hat{1}}$\ \`{a} direita de $(%
\hat{a}^{\dagger })^{n}$ e tamb\'{e}m \`{a} direita de $(b^{\dagger })^{m}.$
Nos c\'{a}lculos por\'{e}m, o operador $\mathbf{\hat{1}}$ funciona como um
dos muitos equivalentes dele: $\hat{S}_{ab}^{\dagger }\hat{S}_{ab}=\mathbf{%
\hat{1}}.$ Lembrando que $\hat{S}_{ab}=\exp [-i\lambda \tau (\hat{a}\hat{b}%
^{\dagger }+\hat{a}^{\dagger }\hat{b})]$ isso acarreta $\hat{S}%
_{ab}|0\rangle _{a}=|0\rangle _{a}$ bem como $\ \hat{S}_{ab}|0\rangle
_{b}=|0\rangle _{b}$ e usando tamb\'{e}m que $\hat{S}_{ab}(\hat{a}^{\dagger
})^{n}\hat{S}_{ab}^{\dagger }=(T\hat{a}^{\dagger }+iR\hat{b}^{\dagger })^{n}$
e $\hat{S}_{ab}(b^{\dagger })^{m}\hat{S}_{ab}^{\dagger }=(R\hat{a}^{\dagger
}-iT\hat{b}^{\dagger })^{m}$\ para todo $n$\ e $m$ inteiros, obtemos o
seguinte resultado,%
\begin{equation}
\left\vert \psi _{ab}^{out}\right\rangle =\sum\limits_{n,m}\frac{C_{n,m}}{%
\sqrt{n!m!}}(T\hat{a}^{\dagger }+iR\hat{b}^{\dagger })^{n}(Tb^{\dagger }+iR%
\hat{a}^{\dagger })^{m}\left\vert 0\right\rangle _{a}\left\vert
0\right\rangle _{b}.  \label{4f}
\end{equation}

\subsection{Caso 4:}

\begin{figure}[tbh]
\begin{center}
%\vspace{-0.5cm}
\includegraphics[width=0.50\textwidth]{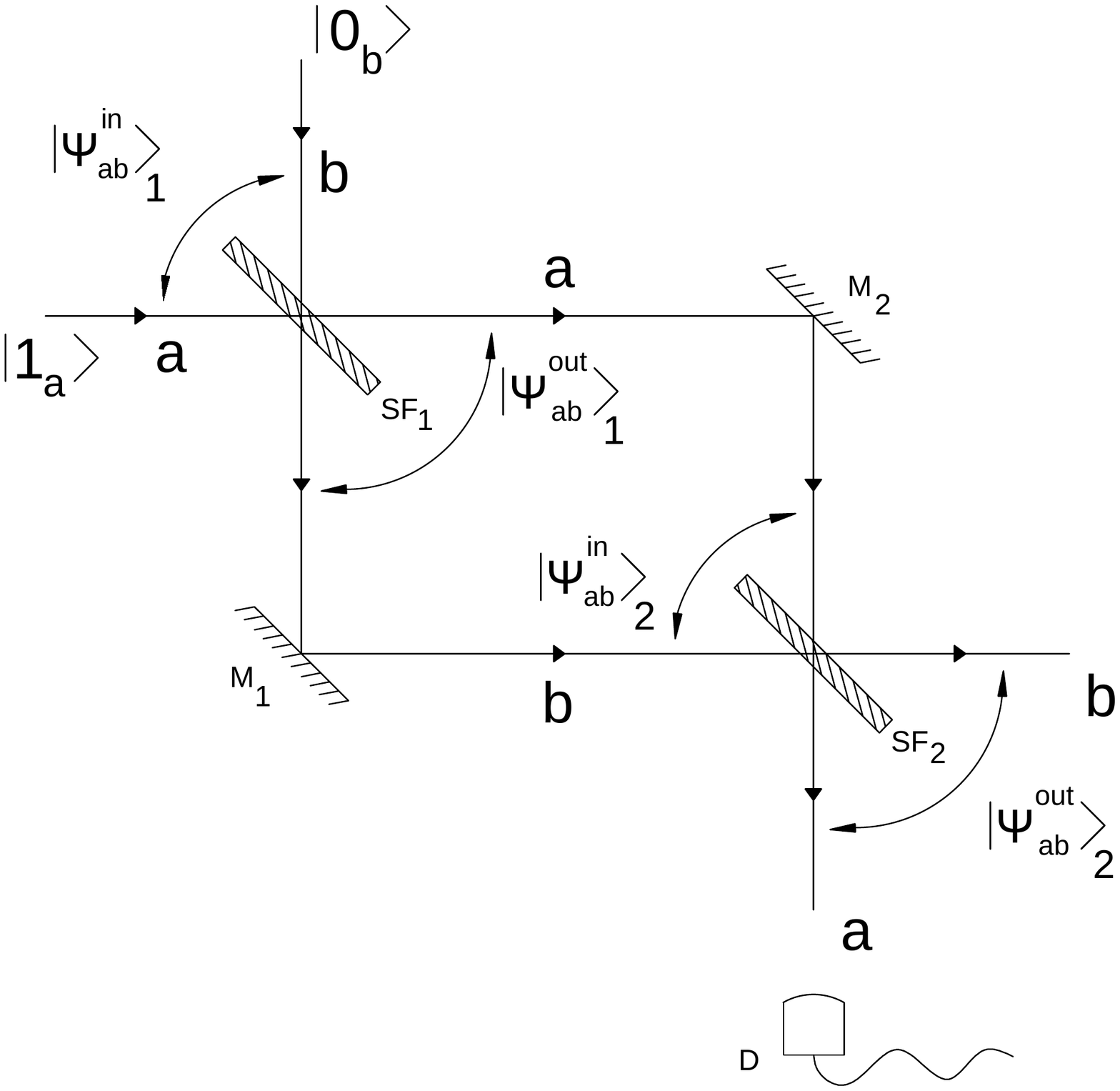}
\end{center}
\par
%\vspace{-1cm}
\caption{Arranjo experimental usando dois espelhos $\mathbf{M}_{\mathbf{1}}$
e $\mathbf{M}_{\mathbf{2}}$, dois separadores de feixes (Beam Splitters) $%
\mathbf{SF}_{\mathbf{1}}$ e $\mathbf{SF}_{\mathbf{2}}$ e detector de f\'{o}%
ton $D$ no modo $\mathbf{a}$ emergente, na vertical. }
\label{bs2}
\end{figure}

Uma aplica\c{c}\~{a}o do resultado acima na \'{O}ptica Qu\^{a}ntica est\'{a}
esbo\c{c}ada na Fig.(\ref{bs2}). Mostra dois feixes do campo luminoso
quantizado incidindo nos modos $\mathbf{a}$ e $\mathbf{b}$ de um $SF$. O
feixe \textit{total} de luz incidente \'{e} representado pelo estado $|\psi
_{ab}^{in}\rangle _{1}=\left\vert 1\right\rangle _{a}\left\vert
0\right\rangle _{b}$ e o feixe \textit{total} emergente \'{e} representado
pelo estado$\ |\psi _{ab}^{out}\rangle _{1}=\hat{S}_{1}|\psi
_{ab}^{in}\rangle _{1}.$Uma similar representa\c{c}\~{a}o se dar\'{a} no
segundo separador de feixes, o $SF_{2}$ da Fig.(\ref{bs2}), para os estados $%
|\psi _{ab}^{in}\rangle _{2}$ e $|\psi _{ab}^{out}\rangle _{2}$ - conforme
detalharemos posteriormente.

%________________________FIGURA 43.1 _________________________________________

Agora, particularizando: na parte superior da Fig.(\ref{bs2}) \'{e} mostrado
no $SF_{1}$ um campo luminoso no estado de $1$ f\'{o}ton, representado pelo s%
\'{\i}mbolo\ $|1\rangle _{a}$, incidente no modo $\mathbf{a,}$ e um campo no
estado de v\'{a}cuo, representado pelo s\'{\i}mbolo\ $|0\rangle _{b}$,
incidente no modo $\mathbf{b}$. Denotamos por $|\psi _{ab}^{in}\rangle _{1}$
o estado\textbf{\ }incidente\ \textit{total} nos modos $\mathbf{a}$ e $%
\mathbf{b}$ do $SF_{1}$ e denotamos por $|\psi _{ab}^{out}\rangle _{1}$ o
estado emergente\ \textit{total} a determinar, nos mesmos dois modos $%
\mathbf{a}$ e $\mathbf{b}$ do $SF_{1}.$ O estado emergente \'{e} resultante
da a\c{c}\~{a}o do operador de evolu\c{c}\~{a}o $\hat{S}_{1}=$ $e^{-i\tau 
\hat{H}/\hbar }$ sobre o estado inicial \textit{total}: $\ \hat{S}_{1}|\psi
_{ab}^{in}\rangle _{1}$\ $\rightarrow |\psi _{ab}^{out}\rangle _{1};$ $\hat{H%
}=\hslash \lambda (\hat{a}^{\dagger }\hat{b}+\hat{a}\hat{b}^{\dagger })$ 
\'{e} o operador hamiltoniano de intera\c{c}\~{a}o comum aos dois $SF_{s}$; $%
\ \hat{a}^{\dagger }(\hat{a})$ e $\hat{b}^{\dagger }(\hat{b})$ s\~{a}o
operadores de cria\c{c}\~{a}o (aniquila\c{c}\~{a}o) para os modos $\mathbf{a}
$ e $\mathbf{b}$; $\tau $\ \'{e} o tempo de travessia nos $SF_{s}$. Temos, 
\begin{equation}
|\psi _{ab}^{out}\rangle _{1}=\hat{S}_{1}|\psi _{ab}^{in}\rangle _{1}.
\label{3c}
\end{equation}%
\qquad

Os respectivos coeficientes de transmiss\~{a}o e reflex\~{a}o, $T_{1}$ e $%
R_{1}$ do $SF_{1},$ ter\~{a}o papel importante nos resultados, conforme
veremos. Substituindo na Eq. (\ref{3c}) o estado inicial $|\psi
_{ab}^{in}\rangle _{1}=\left\vert 1\right\rangle _{a}\left\vert
0\right\rangle _{b}=\hat{a}^{\dagger }\left\vert 0\right\rangle
_{a}\left\vert 0\right\rangle $ mostrado na Fig.(\ref{bs2}), encontramos,
passo a passo o estado na sa\'{\i}da do $SF_{1}$, descrito como:

\begin{eqnarray}
|\psi _{ab}^{out}\rangle _{1} &=&\hat{S}_{1}|\psi _{ab}^{in}\rangle  \notag
\\
&=&\hat{S}_{1}\left\vert 1\right\rangle _{a}\left\vert 0\right\rangle _{b} 
\notag \\
&=&\hat{S}_{1}(\hat{a}^{\dagger }\left\vert 0\right\rangle _{a})\left\vert
0\right\rangle _{b}  \notag \\
&=&\hat{S}_{1}\hat{a}^{\dagger }(\hat{S}_{1}^{\dagger }\hat{S}%
_{1})\left\vert 0\right\rangle _{a}\left\vert 0\right\rangle _{b}  \notag \\
&=&(\hat{S}_{1}\hat{a}^{\dagger }\hat{S}_{1}^{\dagger })\hat{S}%
_{1}\left\vert 0\right\rangle _{a}\left\vert 0\right\rangle _{b}  \notag \\
&=&(T_{1}a^{\dagger }+iR_{1}b^{\dagger })\left\vert 0\right\rangle
_{a}\left\vert 0\right\rangle _{b},  \label{3ca}
\end{eqnarray}%
onde usamos que $\left\vert 1\right\rangle _{a}=$ $\hat{a}^{\dagger
}\left\vert 0\right\rangle _{a}$, inserimos $\hat{S}_{1}^{\dagger }\hat{S}%
_{1}\mathbf{=\hat{1}}$\ \`{a} direita de $\hat{a}^{\dagger }$ e ainda usamos 
$\hat{S}_{1}\left\vert 0\right\rangle _{a}\left\vert 0\right\rangle
_{b}=\left\vert 0\right\rangle _{a}\left\vert 0\right\rangle _{b}$.\ Em
seguida, conforme Fig.(\ref{bs2}), considerando que o estado emergente no $%
SF_{1}$ vai funcionar como estado incidente no $SF_{2},$ isto \'{e}: $|\psi
_{ab}^{in}\rangle _{2}=|\psi _{ab}^{out}\rangle _{1}$\ e usando $\chi
_{2}=\lambda _{2}\tau $ e tamb\'{e}m o operador $\hat{S}_{2}$ em vez de $%
\hat{S}_{1}$ na evolu\c{c}\~{a}o do estado no $SF_{2}$ obtemos, em analogia
com a dedu\c{c}\~{a}o da Eq. (\ref{3ca}),

\begin{eqnarray}
|\psi _{ab}^{out}\rangle _{2} &=&\hat{S}_{2}|\psi _{ab}^{in}\rangle _{2}=%
\hat{S}_{2}|\psi _{ab}^{out}\rangle _{1}  \notag \\
&=&\hat{S}_{2}(T_{1}\hat{a}^{\dagger }+iR_{1}\hat{b}^{\dagger })\left\vert
0\right\rangle _{a}\left\vert 0\right\rangle _{b}  \notag \\
&=&T_{1}(\hat{S}_{2}\hat{a}^{\dagger }\hat{S}_{2}^{\dagger })\left\vert
0\right\rangle _{a}\left\vert 0\right\rangle _{b}+iR_{1}(\hat{S}_{2}\hat{b}%
^{\dagger }\hat{S}_{2}^{\dagger })\left\vert 0\right\rangle _{a}\left\vert
0\right\rangle _{b}  \notag \\
&=&(T_{1}T_{2}-R_{1}R_{2})\left\vert 1\right\rangle _{a}\left\vert
0\right\rangle _{b}  \notag \\
&&+i(R_{1}T_{2}+R_{2}T_{1})\left\vert 0\right\rangle _{a}\left\vert
1\right\rangle _{b},  \label{j1}
\end{eqnarray}%
que resulta em $|\psi _{ab}^{out}\rangle _{2}$ $=$ $|0\rangle _{a}|1\rangle
_{b}$ se $R_{1}=T_{1},$ $R_{2}=T_{2}$ ou se $R_{1}=T_{2},$ $R_{2}=T_{1}.$
Esses dois casos mostram que n\~{a}o emergem f\'{o}tons no modo $\mathbf{a}$
do $SF_{2}$ (interfer\^{e}ncia destrutiva), em acordo com a experi\^{e}ncia.
Agora, se $R_{1}=R_{2}=\sin (\theta ),$ $T_{1}=T_{2}=\cos (\theta )$, a Eq. (%
\ref{j1}) \'{e} obtida na forma 
\begin{equation}
|\psi _{ab}^{out}\rangle _{2}=\cos (2\theta )\left\vert 1\right\rangle
_{a}\left\vert 0\right\rangle _{b}+i\sin (2\theta )\left\vert 0\right\rangle
_{a}\left\vert 1\right\rangle _{b}.  \label{jj}
\end{equation}

Para $\theta =\frac{\pi }{8}$\ o feixe emerge no estado emaranhado com
componentes de mesmo peso, 
\begin{equation}
|\psi _{ab}^{out}\rangle _{2}=\frac{1}{\sqrt{2}}(\left\vert 1\right\rangle
_{a}\left\vert 0\right\rangle _{b}+i\left\vert 0\right\rangle _{a}\left\vert
1\right\rangle _{b});  \label{jj'}
\end{equation}%
para $\theta =\frac{\pi }{4}$ temos o estado emergente $|\psi
_{ab}^{out}\rangle _{2}=\left\vert 0\right\rangle _{a}\left\vert
1\right\rangle _{b}$ e para $\theta =\frac{\pi }{2}$ emerge o mesmo da
entrada: $|\psi _{ab}^{out}\rangle _{2}=\left\vert 1\right\rangle
_{a}\left\vert 0\right\rangle _{b}.$\emph{\ }Agora, se colocam um defasador
em um dos bra\c{c}os do arranjo, retardamos os f\'{o}tons nesse bra\c{c}o e
desaparece a interfer\^{e}ncia destrutiva no detector, esteja este no modo $%
\mathbf{a}$ ou $\mathbf{b}$. Quando esse efeito ocorre sem os defasadores, 
\textbf{significa a presen\c{c}a de espi\~{o}es na rede} - que s\~{a}o
detectados sem que percebam. Isso nos lembra a detec\c{c}\~{a}o `\textit{%
free-interacton measurement}'\ \cite{free}.

O estado de v\'{a}cuo $|0\rangle $ n\~{a}o tem lugar na \'{O}ptica Cl\'{a}%
ssica: no caso usual de estados com muitos f\'{o}tons, ele n\~{a}o tem papel
relevante em geral. Mas, para estados com poucos f\'{o}tons, tipo $1,2,3$ f%
\'{o}tons,... o estado $|0\rangle $ torna-se muito importante; \'{e} quando
os experimentos discordam do resultado te\'{o}rico cl\'{a}ssico. A inibi\c{c}%
\~{a}o de fotocontagem nesse arranjo \'{o}ptico n\~{a}o ocorre no tratamento
qu\^{a}ntico usando estados `\textit{cl\'{a}ssicos}', tipo estado t\'{e}%
rmico ou mesmo o estado coerente - o mais cl\'{a}ssico dentre os estados\ n%
\~{a}o cl\'{a}ssicos. A prop\'{o}sito, o estado coerente $\left\vert \alpha
\right\rangle $\ \'{e} definido no tratamento qu\^{a}ntico e ele est\'{a} na
fronteira entre os estados cl\'{a}ssicos e os estados qu\^{a}nticos.

\subsection{Caso 5:}

\begin{figure}[tbh]
\begin{center}
%\vspace{-0.5cm}
\includegraphics[width=0.50\textwidth]{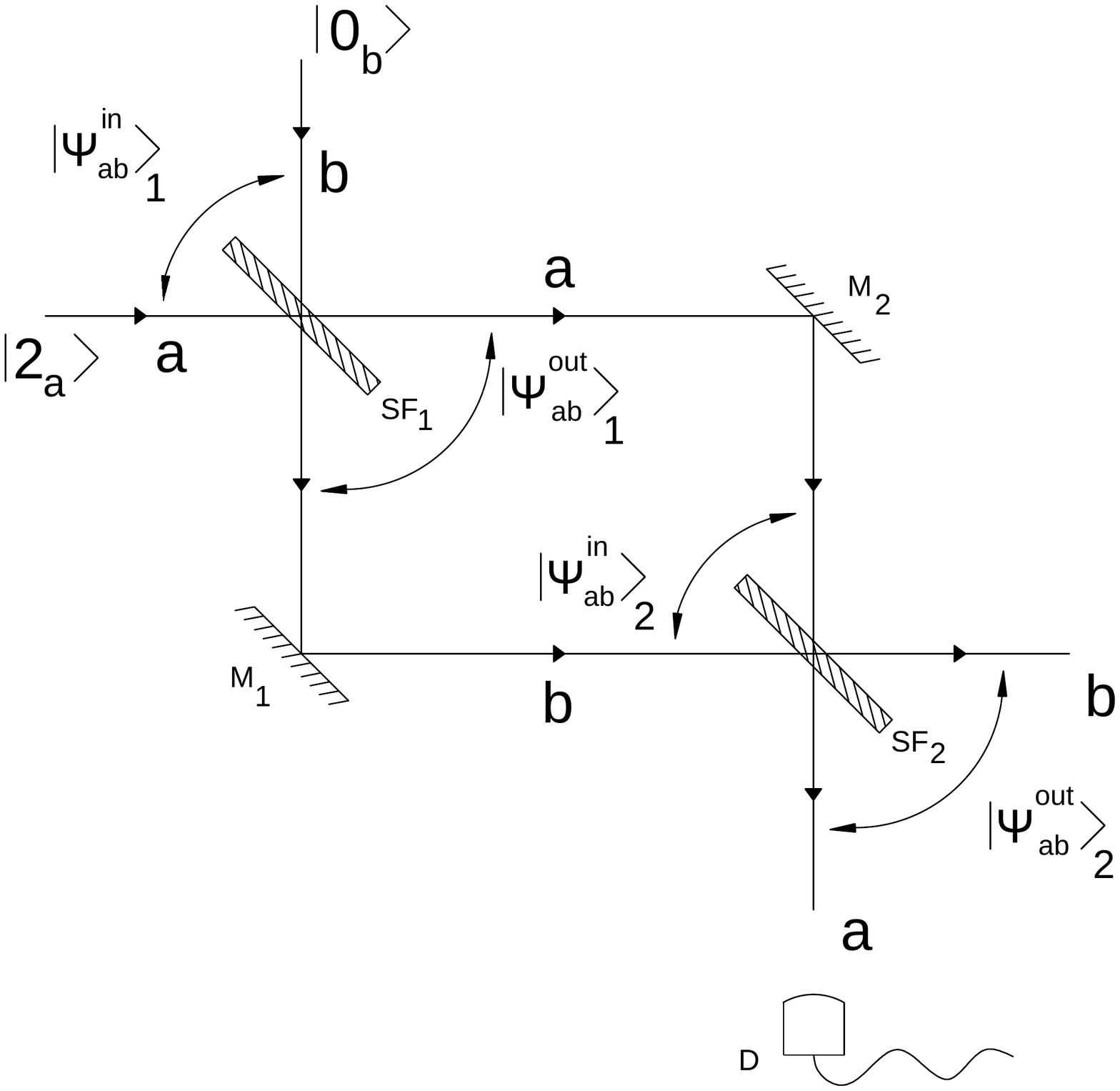}
\end{center}
\par
%\vspace{-1cm}
\caption{O mesmo que na Fig.(\protect\ref{bs2}), com $|\protect\psi %
_{ab}^{in}\rangle _{1}=|2\rangle _{a}\left\vert 0\right\rangle _{b}.$}
\label{bs5}
\end{figure}

Vamos considerar agora um feixe luminoso no estado $\left\vert
2\right\rangle ,$ de dois f\'{o}tons, incidindo no modo $\mathbf{a}$\textbf{%
\ }do\textbf{\ }$SF_{1}$ e outro feixe no estado de v\'{a}cuo, $\left\vert
0\right\rangle ,$ incidindo no bra\c{c}o $\mathbf{b,}$\ do mesmo $SF_{1},$
ver Fig.(\ref{bs5}). Esse estado inicial \textit{total} no $SF_{1}$ ser\'{a}
representado matematicamente por,

\begin{equation}
|\psi _{ab}^{in}\rangle _{1}=|2\rangle _{a}\left\vert 0\right\rangle _{b}.
\label{4y}
\end{equation}

%________________________FIGURA 43.1 _________________________________________

Seguindo o procedimento anterior, que levou \`{a} Eq. (\ref{j1}),
encontramos na sa\'{\i}da do $SF_{2}$ este estado emergente,%
\begin{widetext}
\begin{eqnarray}
|\psi _{ab}^{out}\rangle _{2} &=&(T_{1}T_{2}-R_{1}R_{2})^{2}\left\vert
2\right\rangle _{a}\left\vert 0\right\rangle
_{b}-(T_{1}R_{2}+T_{2}R_{1})^{2}\left\vert 0\right\rangle _{a}\left\vert
2\right\rangle _{b}  \notag \\
&&+i\sqrt{2}%
(T_{1}^{2}T_{2}R_{2}+T_{1}T_{2}^{2}R_{1}-T_{1}R_{1}R_{2}^{2}-R_{1}^{2}R_{2}T_{2})\left\vert 1\right\rangle _{a}\left\vert 1\right\rangle _{b}.
\label{j2}
\end{eqnarray}%
Neste caso, para $R_{1}=R_{2}=\sin (\theta )$ e $T_{1}=T_{2}=\cos (\theta )$
a Eq.(\ref{j2}) resulta na forma,
\begin{equation}
|\psi _{ab}^{out}\rangle _{2}=\cos ^{2}(2\theta )\left\vert 2\right\rangle
_{a}\left\vert 0\right\rangle _{b}-\sin ^{2}(2\theta )\left\vert
0\right\rangle _{a}\left\vert 2\right\rangle _{b}+\frac{i}{\sqrt{2}}\sin
(4\theta )\left\vert 1\right\rangle _{a}\left\vert 1\right\rangle _{b},
\label{j3}
\end{equation}

\end{widetext}a qual, para valores de $R_{1},$ $R_{2},T_{1}\ $e$\ T_{2}$ que
a qual, para os valores de $\theta =0,\frac{\pi }{2}~e~\pi $, fornece o
estado emergindo do $SF_{2},$ 
\begin{equation}
|\psi _{ab}^{out}\rangle _{2}=\left\vert 2\right\rangle _{a}\left\vert
0\right\rangle _{b},  \label{j4}
\end{equation}%
que coincide com o estado incidente no $SF_{1}.$ O caso $\theta =\frac{\pi }{%
4}$ mostra troca de estados nos modos $\mathbf{a}$ e $\mathbf{b}$, 
\begin{equation}
|\psi _{ab}^{out}\rangle _{2}=\left\vert 0\right\rangle _{a}\left\vert
2\right\rangle _{b},  \label{j5}
\end{equation}%
enquanto que, para $\theta =\frac{\pi }{8}$ o estado emergente \'{e} este,

\begin{equation}
|\psi _{ab}^{out}\rangle _{2}=\frac{1}{2}(\left\vert 2\right\rangle
_{a}\left\vert 0\right\rangle _{b}-\left\vert 0\right\rangle _{a}\left\vert
2\right\rangle _{b}+i\sqrt{2}\left\vert 1\right\rangle _{a}\left\vert
1\right\rangle _{b}).  \label{j6}
\end{equation}

\subsection{Caso 6:}

\begin{figure}[tbh]
\begin{center}
%\vspace{-0.5cm}
\includegraphics[width=0.50\textwidth]{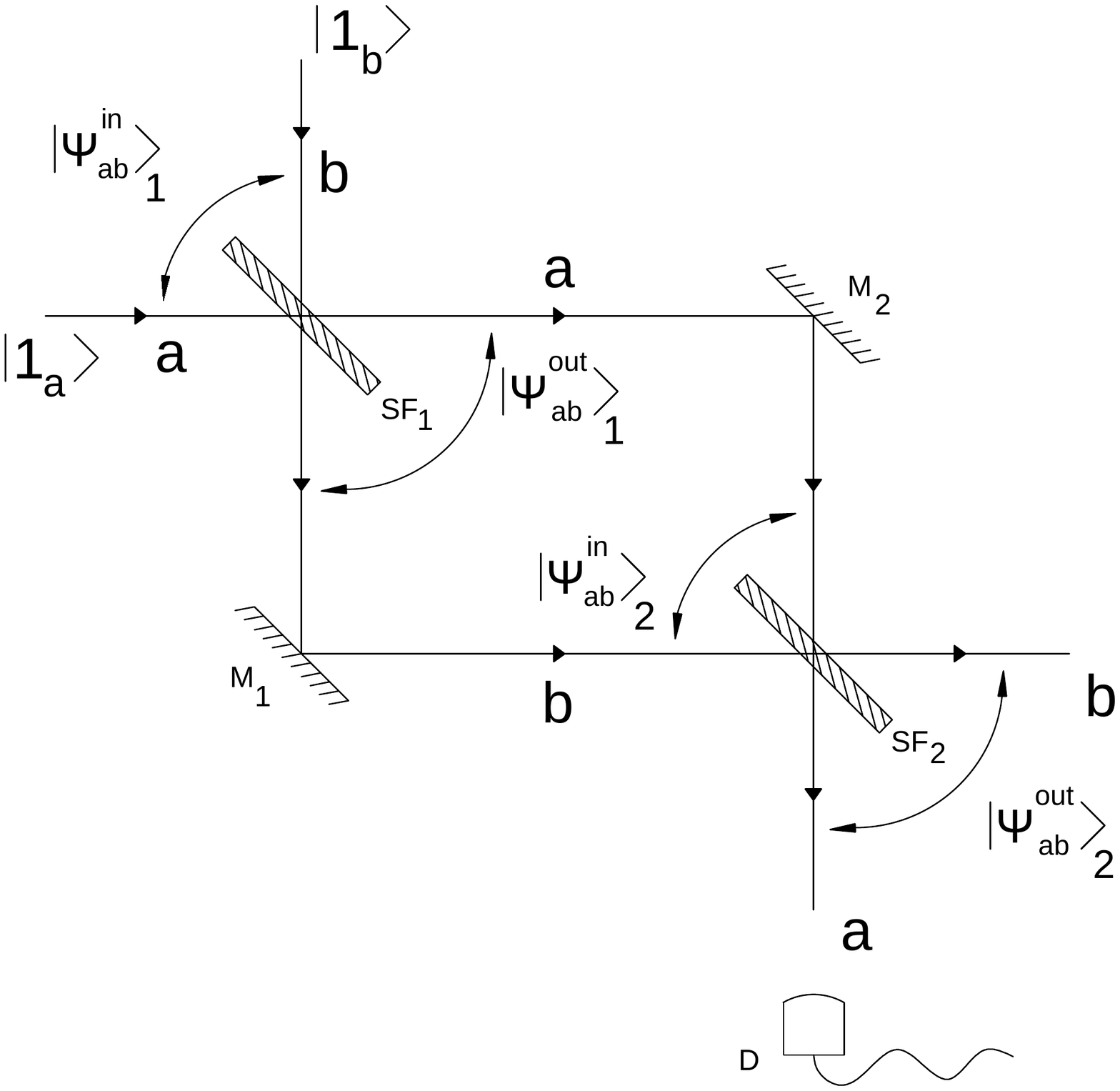}
\end{center}
\par
%\vspace{-1cm}
\caption{O mesmo que na Fig.(\protect\ref{bs2}), com $|\protect\psi %
_{ab}^{in}\rangle _{1}=|1\rangle _{a}\left\vert 1\right\rangle _{b}.$}
\label{bs6}
\end{figure}

Neste caso assumiremos os dois feixes luminosos no estado de $1$ f\'{o}ton, $%
\left\vert 1\right\rangle _{a}$ e $\left\vert 1\right\rangle _{b}$,
incidindo nos bra\c{c}os $\mathbf{a}$\textbf{\ e }$\mathbf{b}$ do $SF_{1},$
ver Fig.(\ref{bs6}). O estado inicial \textit{total} incidente no primeiro $%
SF_{1}$ \'{e} representado por,%
%________________________FIGURA 43.1 _________________________________________

\begin{equation}
|\psi _{ab}^{in}\rangle _{1}=|1\rangle _{a}\left\vert 1\right\rangle _{b}.
\label{j8}
\end{equation}

Utilizando os mesmos procedimentos matem\'{a}ticos anteriores, isto \'{e},
tratando o estado \textit{total} emergente do $SF_{1}$ como o estado \textit{%
total} incidente no $SF_{2},$ encontramos o estado \textit{total} emergente
na sa\'{\i}da deste, como:%
\begin{widetext}
\begin{eqnarray}
|\psi _{ab}^{out}\rangle _{2}
&=&[(T_{1}T_{2}-R_{1}R_{2})^{2}-(T_{1}R_{2}+T_{2}R_{1})^{2}]\left\vert
1\right\rangle _{a}\left\vert 1\right\rangle _{b}  \notag \\
&&+i\sqrt{2}%
(T_{1}^{2}T_{2}R_{2}+T_{1}T_{2}^{2}R_{1}-T_{1}R_{1}R_{2}^{2}-R_{1}^{2}R_{2}T_{2})[\left\vert 2\right\rangle _{a}\left\vert 0\right\rangle _{b}+\left\vert 0\right\rangle _{a}\left\vert 2\right\rangle _{b}]%
\text{ }.  \label{j9}
\end{eqnarray}%
Agora, no o caso $R_{1}=R_{2}=\sin (\theta ),$ $T_{1}=T_{2}=\cos (\theta )$,
a Eq.(\ref{j9}) \'{e} escrita como,%
\begin{equation}
|\psi _{ab}^{out}\rangle _{2}=\cos (4\theta )\left\vert 1\right\rangle
_{a}\left\vert 1\right\rangle _{b}+\frac{i}{\sqrt{2}}\sin (4\theta
)(\left\vert 2\right\rangle _{a}\left\vert 0\right\rangle _{b}+\left\vert
0\right\rangle _{a}\left\vert 2\right\rangle _{b}).  \label{j10}
\end{equation}%
\end{widetext}Para $\theta =0,$ $\frac{\pi }{2},\frac{\pi }{4}$, e $\pi $\ o
estado emergente do $SF_{2}$ coincide com o estado de entrada no $SF_{1},$ 
\begin{equation}
|\psi _{ab}^{out}\rangle _{2}=\left\vert 1\right\rangle _{a}\left\vert
1\right\rangle _{b},  \label{j11}
\end{equation}%
enquanto que para $\theta =\frac{\pi }{8}$ o estado emergente no $SF_{2}$
exibe componentes de $2$ f\'{o}tons nos modos $\mathbf{a}$ ou $\mathbf{b}$,
constituindo-se em um m\'{e}todo para gerar o estado de Fock $\left\vert
2\right\rangle $ em um modo do campo viajante a partir dos estados $%
\left\vert 1\right\rangle $ e $\left\vert 0\right\rangle ,$ 
\begin{equation}
|\psi _{ab}^{out}\rangle _{2}=\frac{i}{\sqrt{2}}(\left\vert 2\right\rangle
_{a}\left\vert 0\right\rangle _{b}+\left\vert 0\right\rangle _{a}\left\vert
2\right\rangle _{b}).  \label{j13}
\end{equation}

Este resultado mostra que, se detectamos o estado acima no modo \textbf{a}
em $\left\vert 0\right\rangle _{a}$, isto deixa o feixe do modo \textbf{b}
no estado $\left\vert 2\right\rangle _{b}$\textbf{\ }e, se ele\textbf{\ }%
incidir em um $SF_{3},$ em cujo outro modo incide feixe no estado $%
\left\vert 0\right\rangle $ ou \ $\left\vert 1\right\rangle $ , geramos na
saida deste\textbf{\ }$SF_{3}$ os estados de Fock $\left\vert 3\right\rangle 
$ e $\left\vert 4\right\rangle ,$ respectivamente, pela conveniente escolha
de $\theta .$ N\~{a}o podemos por\'{e}m esquecer que a probabilidade de detec%
\c{c}\~{a}o de estados de Fock assim obtidos, decaem com a pot\^{e}ncia $%
\frac{1}{2^{n}}$, onde $n$ \'{e} a ordem do $SF.$

\subsection{Caso 7:}

\begin{figure}[tbh]
\begin{center}
%\vspace{-0.5cm}
\includegraphics[width=0.50\textwidth]{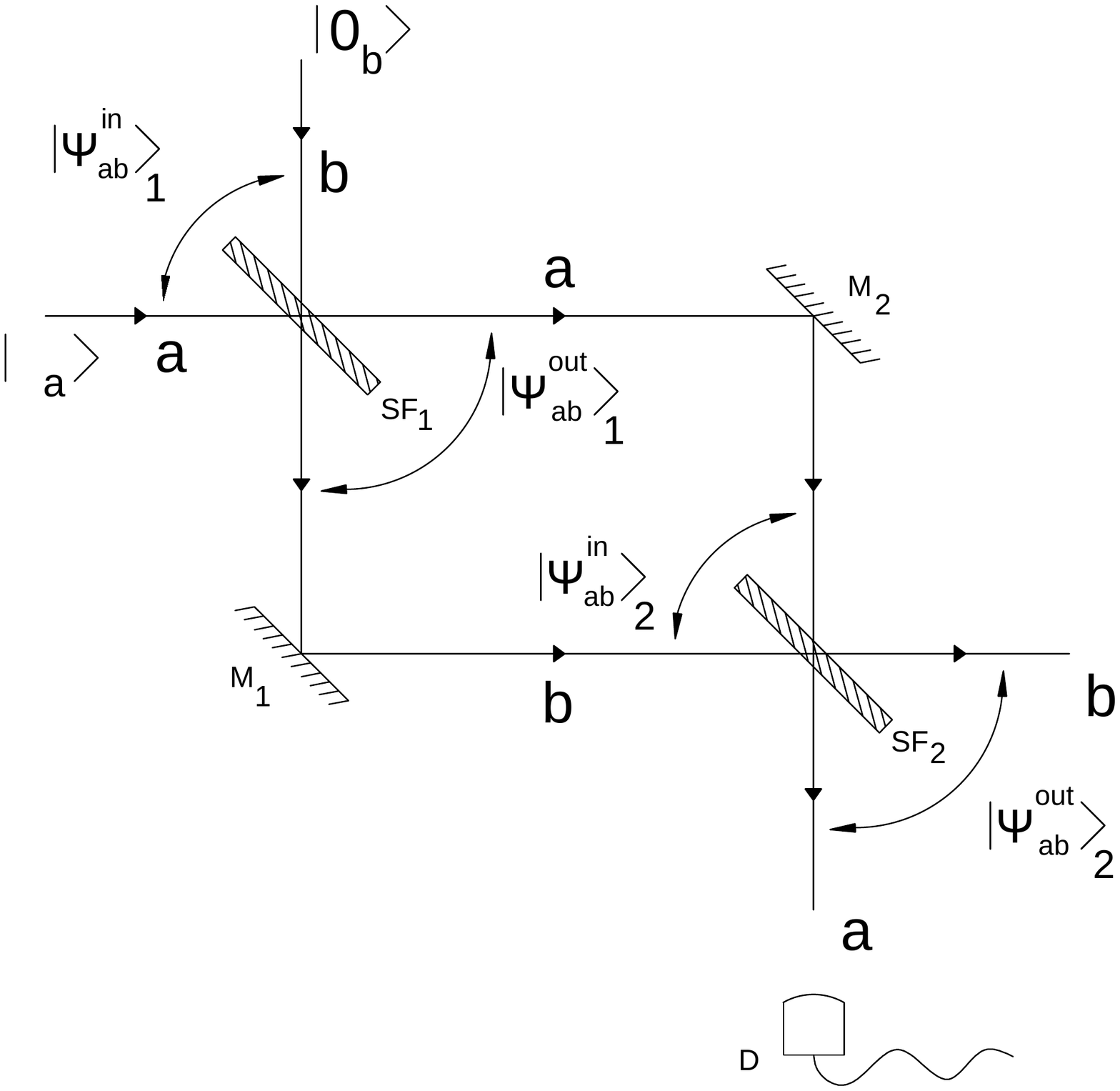}
\end{center}
\par
%\vspace{-1cm}
\caption{O mesmo que na Fig.(\protect\ref{bs2}), com $|\protect\psi %
_{ab}^{in}\rangle _{1}=|\protect\alpha \rangle _{a}\left\vert 0\right\rangle
_{b}.$}
\label{bs7}
\end{figure}

Vamos considerar um feixe de luz no estado coerente $\left\vert \alpha
\right\rangle $ incidindo pelo bra\c{c}o $\mathbf{a}$ do $SF_{1}$\ e outro
feixe no estado de v\'{a}cuo, $\left\vert 0\right\rangle ,$ no bra\c{c}o $%
\mathbf{b,}$\ ver Fig.(\ref{bs7}); o estado inicial \textit{total} incidente
no $SF_{1}$ \'{e} expresso por,

\begin{equation}
|\psi _{ab}^{in}\rangle _{1}=|\alpha \rangle _{a}\left\vert 0\right\rangle
_{b},  \label{j14}
\end{equation}

%________________________FIGURA 43.1 _________________________________________
e na sa\'{\i}da do $SF_{2}$ temos este estado do feixe emergente \textit{%
total}:%
\begin{widetext}
\begin{eqnarray}
|\psi _{ab}^{out}\rangle _{2} &=&e^{[\alpha (T_{1}T_{2}-R_{1}R_{2})\hat{a}%
^{\dagger }-\alpha ^{\ast }(T_{1}T_{2}-R_{1}R_{2})\hat{a}]}e^{[\alpha
(iT_{1}R_{2}+iR_{1}T_{2})\hat{b}^{\dagger }+\alpha ^{\ast
}(iT_{1}R_{2}+iR_{1}T_{2})\hat{b}]}\left\vert 0\right\rangle _{a}\left\vert
0\right\rangle _{b},  \notag \\
&=&\left\vert (T_{1}T_{2}-R_{1}R_{2})\alpha \right\rangle _{a}\left\vert
(T_{1}R_{2}+R_{1}T_{2})i\alpha \right\rangle _{b}.  \label{4f1}
\end{eqnarray}%
Considerando o caso particular $R_{1}=R_{2}=\sin (\theta )$ e $%
T_{1}=T_{2}=\cos (\theta )$ temos,%
\begin{equation*}
|\psi _{ab}^{out}\rangle _{2}=\left\vert \cos (2\theta )\alpha \right\rangle
_{a}\left\vert i\sin (2\theta )\alpha \right\rangle _{b};
\end{equation*}%
\end{widetext}para $\theta =0,$ $\frac{\pi }{2}$ e $\pi $ o estado na sa%
\'{\i}da do $SF_{2}$ coincide com o estado de entrada, 
\begin{equation}
|\psi _{ab}^{out}\rangle _{2}=\left\vert \alpha \right\rangle _{a}\left\vert
0\right\rangle _{b}.  \label{4f2}
\end{equation}%
Para $\theta =\frac{\pi }{4}$ o estado emergente mostra uma quase troca de
estados,

\begin{equation}
|\psi _{ab}^{out}\rangle _{2}=\left\vert 0\right\rangle _{a}\left\vert
i\alpha \right\rangle _{b},  \label{4f3}
\end{equation}%
havendo de fato uma troca de estat\'{\i}stica, pois $|i\alpha \rangle =|e^{i%
\frac{\pi }{2}}\alpha \rangle $ tamb\'{e}m \'{e} um (outro) estado coerente,
resultado de girar $|\alpha \rangle $ de $\frac{\pi }{2}$ no espa\c{c}o de
fase da Mec\^{a}nica Qu\^{a}ntica; j\'{a} o valor $i\alpha $ representa um
giro de $\frac{\pi }{2}$ no valor $\alpha ,$\ no plano complexo. Estados
coerentes t\^{e}m mesma estat\'{\i}stica, de Poisson.

Para $\theta =\frac{\pi }{8}$ temos,%
\begin{equation}
|\psi _{ab}^{out}\rangle _{2}=\left\vert \frac{\alpha }{\sqrt{2}}%
\right\rangle _{a}\left\vert i\frac{\alpha }{\sqrt{2}}\right\rangle _{b}.
\label{4f5}
\end{equation}%
significando dois estados coerentes, um em cada modo do $SF_{2},$ com
conserva\c{c}\~{a}o do n\'{u}mero de f\'{o}tons: $\langle \hat{n}\rangle
_{SF_{1}}=\left\vert \alpha \right\vert _{a}^{2}+|0|^{2}=\left\vert \alpha
\right\vert _{a}^{2}$ que coincide com $\langle \hat{n}\rangle _{SF_{2}}=%
\frac{1}{2}(|\alpha |_{a}^{2}+|i\alpha |_{b}^{2})=\left\vert \alpha
\right\vert _{a}^{2}.$ Tinha de ser assim pois o operador de evolu\c{c}\~{a}%
o, $\hat{S}_{ab},$ \'{e} unit\'{a}rio.

\subsection{Caso 8:}

\begin{figure}[tbh]
\begin{center}
%\vspace{-0.5cm}
\includegraphics[width=0.50\textwidth]{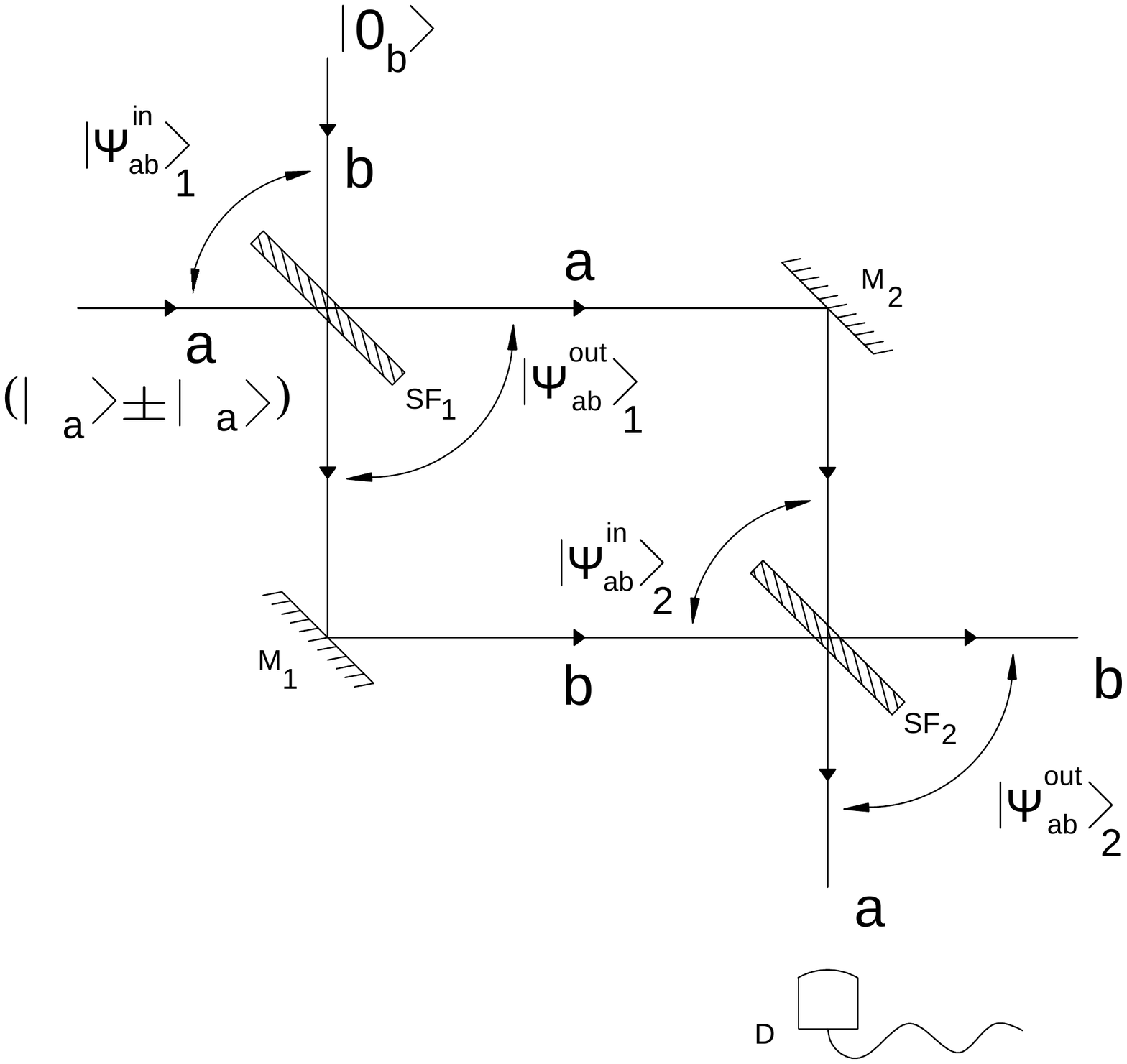}
\end{center}
\par
%\vspace{-1cm}
\caption{O mesmo que na Fig.(\protect\ref{bs2}), com $|\protect\psi %
_{ab}^{in}\rangle _{1}=\protect\eta (\left\vert \protect\alpha \right\rangle
_{a}\pm \left\vert \protect\beta \right\rangle _{a})\left\vert
0\right\rangle _{b}.$}
\label{bs8}
\end{figure}

Vamos considerar um feixe de luz no estado de superposi\c{c}\~{a}o $\eta
(\left\vert \alpha \right\rangle _{a}\pm \left\vert \beta \right\rangle
_{a}),$ formado de dois estados coerentes, $\left\vert \alpha \right\rangle
_{a}$ e $\left\vert \beta \right\rangle _{a}$ , incidente no bra\c{c}o $%
\mathbf{a}$\textbf{\ }do $SF_{1}$; consideremos ainda o estado de v\'{a}cuo
de f\'{o}ton, $\left\vert 0\right\rangle ,$ no bra\c{c}o $\mathbf{b}$\ do
mesmo $SF_{1},$\ ver Fig.(\ref{bs8}); o par\^{a}metro $\eta =$ $(|\alpha
|^{2}$ $+$ $|\beta |^{2}+$\ $2Re$ $\langle \alpha |\beta \rangle )^{-\frac{1%
}{2}}$\ \'{e} o fator de normaliza\c{c}\~{a}o do estado superposto. O estado
inicial \textit{total} incidindo no $SF_{1}$ \'{e} escrito como,

\begin{equation}
|\psi _{ab}^{in}\rangle _{1}=\eta (\left\vert \alpha \right\rangle _{a}\pm
\left\vert \beta \right\rangle _{a})\left\vert 0\right\rangle _{b}.
\label{4f6}
\end{equation}

%________________________FIGURA 43.1 _________________________________________

Feitas as contas, obtemos o estado \textit{total} emergindo do $SF_{2}$ dado
por,%
\begin{widetext}
\begin{equation}
|\psi _{ab}^{out}\rangle _{2}=\eta (\left\vert (T_{1}T_{2}-R_{1}R_{2})\alpha
\right\rangle _{a}\left\vert (T_{1}R_{2}+R_{1}T_{2})i\alpha \right\rangle
_{b}\pm \left\vert (T_{1}T_{2}-R_{1}R_{2})\beta \right\rangle _{a}\left\vert
(T_{1}R_{2}+R_{1}T_{2})i\beta \right\rangle _{b}).  \label{4f7}
\end{equation}%
No caso particular $R_{1}=R_{2}=\sin (\theta ),$ $T_{1}=T_{2}=\cos (\theta )$
a express\~{a}o acima assume a forma,%
\begin{equation}
|\psi _{ab}^{out}\rangle _{2}=\left\vert (\cos 2\theta )\alpha \right\rangle
_{a}\left\vert i(\sin 2\theta )\alpha \right\rangle _{b}\pm \left\vert (\cos
2\theta )\beta \right\rangle _{a}\left\vert i(\sin 2\theta )\beta
\right\rangle _{b}.  \label{4f8}
\end{equation}%
\end{widetext}Notamos que para $\theta =0,\frac{\pi }{2}~e~\pi $ temos o
estado emergente do $SF_{2},$ 
\begin{equation}
|\psi _{ab}^{out}\rangle _{2}=\eta (\left\vert \alpha \right\rangle _{a}\pm
\left\vert \beta \right\rangle _{a})\left\vert 0\right\rangle _{b},
\label{4f9}
\end{equation}%
que coincide com o estado incidente no $SF_{1}$.

Para $\theta =\frac{\pi }{4}$ o estado superposto transfere-se do modo $%
\mathbf{a}$ para o modo $\mathbf{b}$\emph{,} sofrendo um giro de $\frac{\pi 
}{2}$ nas duas componentes coerentes. 
\begin{equation}
|\psi _{ab}^{out}\rangle _{2}=\eta \left\vert 0\right\rangle _{a}(\left\vert
i\alpha \right\rangle _{b}\pm \left\vert i\beta \right\rangle _{b}),
\label{4f11}
\end{equation}%
enquanto que para $\theta =\frac{\pi }{8}$ temos um estado entrela\c{c}ando
os modos $\mathbf{a}$ e $\mathbf{b}$, 
\begin{equation}
|\psi _{ab}^{out}\rangle _{2}=\eta (\left\vert \alpha \right\rangle
_{a}\left\vert i\alpha \right\rangle _{b}\pm \left\vert \beta \right\rangle
_{a}\left\vert i\beta \right\rangle _{b}).  \label{4f12}
\end{equation}

\section{Conclus\~{a}o}

Neste breve texto apresentamos alguns resultados do uso de estados em modos
viajantes. O arranjo utiliza costumeiros dispositivos \'{o}pticos de
experimentos em laborat\'{o}rios: fontes de feixes luminosos preparados em
convenientes estados (estados de Fock, coerentes, superpostos, ...);
separadores de feixes (beam splitters), espelhos e detectores de f\'{o}tons.
Outros arranjos de laborat\'{o}rios usam dispositivos adicionais, como
prismas, redes de difra\c{c}\~{a}o, fibras \'{o}pticas, etc. \'{E} mostrado
como a \'{a}lgebra de operadores aplicada ao caso de dois feixes incidentes
em um $SF$ fornecendo o estado total do feixe que emerge no $SF$. O
resultado \'{e} estendido a um arranjo tipo \textquotedblleft espectr\^{o}%
metro de Mach Zehnder\textquotedblright\ onde obtemos o estado do feixe
luminoso na saida de um segundo $SF$, o $SF_{2}$, ap\'{o}s usar estado de
saida no $SF_{1}$ como estado de entrada no $SF_{2}.$ Isso \'{e} obtido com
ajuda de dois espelhos, como mostrado na Fig.(\ref{bs2}). \ A aplica\c{c}%
\~{a}o ao caso de v\'{a}rios estados mais simples na entrada do $SF_{1}$ foi
mostrada, com os estados de $\left\vert 1\right\rangle $ e $\left\vert
2\right\rangle $ f\'{o}tons, bem como usando estados coerentes e uma de suas
importantes superposi\c{c}\~{o}es (ver Eq. (\ref{4f6})). Alguns resultados
mostrados s\~{a}o: a troca de estados entre os $2$ modos do arranjo (ver
Eqs. (\ref{j14}) e (\ref{4f3}) ) e a subdivis\~{a}o de um estado coerente
(que incide no $SF_{1}$) em dois estados coerentes emergindo do $SF_{2}$
(ver Eqs. (\ref{j14}) e (\ref{4f5})). O esquema fornece tamb\'{e}m a produ%
\c{c}\~{a}o de estados de Fock mais excitados a partir de dois estados de
Fock menos excitados (ver Eqs. (\ref{j8}) e (\ref{j13})); outro resultado 
\'{e} a produ\c{c}\~{a}o de estados entrela\c{c}ados a partir de superposi%
\c{c}\~{o}es de dois estados coerentes (ver Eqs.\ (\ref{4f6}) e (\ref{4f12}%
)). Outros estados interessantes podem tamb\'{e}m ser obtidos pela escolha
de diferentes valores dos par\^{a}metros $T_{i},$ $R_{i},$ $i=1,2$, bem como
da conveniente escolha do estado detectado em um dos modos da sa\'{\i}da.
Por exemplo, \'{e} de grande interesse na gera\c{c}\~{a}o de estados entrela%
\c{c}ados mistos, de \'{a}tomo e campo, dispor previamente de estados
coerentes e de Fock, entrela\c{c}ados \cite{e6}. Sobre gera\c{c}\~{a}o de
distintos estados entrela\c{c}ados em modos viajantes ou usando modos
viajantes, citamos a Ref. \cite{e4}, que utiliza modos viajantes cl\'{a}%
ssicos para entrela\c{c}ar estados at\^{o}micos, e a Ref. \cite{e8}, que
emprega sofisticado esquema experimental e convenientes estados para entrela%
\c{c}ar dois estados do tipo \textquotedblleft gato de Schr\"{o}%
dinger\textquotedblright . Finalmente, se substituirmos os estados de
entrada no $SF_{1}$, por exemplo os estados $|0\rangle $ e $|1\rangle $ ou
outro par qualquer, pelos estados de polariza\c{c}\~{a}o $|V\rangle $ e $%
|H\rangle ,$ horizontal e vertical, respectivamente, temos um cen\'{a}rio
que facilita procedimentos tomogr\'{a}ficos \cite{hh} permitindo o
reconhecimento e a manipula\c{c}\~{a}o dos estados na sa\'{\i}da do $SF_{2}.$

\section{Agradecimentos}

Os autores agradecem ao CNPq e FAPEG pelo suporte parcial deste trabalho, e
ao Professor Rafael de Morais Gomes (IF/UFG) pela refer\^{e}ncia \cite{hh}.

\end{document}